\documentclass{aa}
\usepackage{txfonts}
\usepackage{epsfig}
\usepackage{subfigure}
\usepackage{multirow}

\usepackage{times}
\usepackage{natbib}
\usepackage{longtable,lscape}
\bibpunct{(}{)}{;}{a}{}{,}

\begin{document}

\title{CAIXA: a Catalogue of AGN In the XMM-\textit{Newton} Archive\\ II. Multiwavelength correlations}
\author{Stefano Bianchi\inst{1,2}, Nuria Fonseca Bonilla\inst{2}, Matteo Guainazzi\inst{2}, Giorgio Matt \inst{1}, Gabriele Ponti \inst{3}}

\offprints{Stefano Bianchi\\ \email{bianchi@fis.uniroma3.it}}

\institute{Dipartimento di Fisica, Universit\`a degli Studi Roma Tre, via della Vasca Navale 84, 00146 Roma, Italy
\and XMM-Newton Science Operations Center, European Space Astronomy Center, ESA, Apartado 50727, E-28080 Madrid, Spain
\and Laboratoire APC, UMR 7164, 10 rue A. Domon et L. Duquet, 75205 Paris, France}

\date{Received / Accepted}

\authorrunning{S. Bianchi et al.}

\abstract
{}
{The availability of large amounts of multiwavelength data allows us to perform an extensive statistical analysis to look for correlations between different parameters of AGN. The physical implications of these correlations, when considered within the framework of current AGN models, can be enlightening in the resolution of the problems and the open issues which still characterise this class of sources.}
{We presented CAIXA, a Catalogue of AGN in the XMM-\textit{Newton} Archive, in a companion paper. It consists of radio-quiet X-ray unobscured sources, which cover a range in X-ray luminosities between L$_\mathrm{2-10\,keV} = 2.0\times 10^{41}$ and $3.9\times10^{46}$ erg s$^{-1}$, and in redshift from z=0.002 to z=4.520. Here, a systematic search for correlations between the X-ray spectral properties and the multiwavelength data was performed for the sources in CAIXA. All the significant ($>99.9\%$ confidence level) correlations are discussed along with their physical implications on current models of AGN.}
{Two main correlations are discussed in this paper: a) a very strong anti-correlation between the FWHM of the H$\beta$ optical line and the ratio between the soft and the hard X-ray luminosity. Although similar anti-correlations between optical line width and X-ray spectral steepness have already been discussed in the literature (see e.g., Laor et al. 1994, Boller et al. 1996, Brandt et al. 1997), we consider the formulation we present in this paper is more fundamental, as it links model-independent quantities. Coupled with a strong anti-correlation between the V to hard X-ray flux ratio and the H$\beta$ FHWM, it supports scenarios for the origin of the soft excess in AGN, which require strong suppression of the hard X-ray emission; b) a strong (and expected) correlation between the X-ray luminosity and the black hole mass. Its slope, flatter than 1, is consistent with Eddington ratio-dependent bolometric corrections, such as that recently proposed by Vasudevan \& Fabian (2009). Moreover, we critically review through various statistical tests the role that distance biases play in the strong radio to X-ray luminosity correlation found in CAIXA and elsewhere; we conclude that only complete, unbiased samples (such as that recently published by Behar \& Laor, 2008) should be used to draw observational constraints on the origin of radio emission in radio-quiet AGN.}
{}


\maketitle

\section{Introduction}

It is now well established that active galactic nuclei (AGN) are powered by accretion onto a central supermassive black hole \citep{lb69}. Different flavours of unification models are quite successful in explaining most of the varied phenomenology of these sources in terms of a fundamental scenario, which explains the characteristic signatures in the spectra of AGN: a thermal disc emitting in the optical and ultraviolet; a corona of hot electrons producing X-ray photons via inverse Compton scattering of the photons from the disc; dusty gas which reprocesses the nuclear emission into infrared emission and into other X-ray spectral components. However, many aspects related to fundamental properties of AGN are still to be fully understood and many competing models have been developed and need to be tested. One fruitful approach to this kind of investigation is to gather as much multiwavelength data as possible on large numbers of sources and perform a systematic statistical analysis looking for correlations between different parameters. Much progress have been done since the first seminal attempts \citep[e.g.][]{bg92}, thanks to the availability of ever-increasing data archives. In particular, in the last few years large amounts of high-quality X-ray data have become easily accessible.

We presented CAIXA, a Catalogue of AGN in the XMM-\textit{Newton} Archive, in a companion paper \citep[][B09 from now on]{bianchi09}. It consists of all the radio-quiet X-ray unobscured ($\mathrm{N_H}<2\times10^{22}$ cm$^{-2}$) Active Galactic Nuclei (AGN) observed by XMM-\textit{Newton} in targeted observations, whose data are public as of March 2007. With its 156 sources, this is the largest catalogue of high signal-to-noise X-ray spectra of AGN. It covers a range in X-ray luminosities between L$_\mathrm{2-10\,keV} = 2.0\times 10^{41}$ and $3.9\times10^{46}$ erg s$^{-1}$, and in redshift from z=0.002 to z=4.520. For each source in CAIXA we have collected observable and physical parameters more directly related to the accretion onto the supermassive black hole (BH), and less affected by contamination by the host galaxy. Therefore, the EPIC pn X-ray data are complemented by multiwavelength data found in the literature: BH masses, Full Width Half Maximum (FWHM) of H$\beta$, radio and optical fluxes.

We refer the reader to B09 for all the details on the selection of the catalogue, the X-ray data reduction analysis and the coverage of the multiwavelength data. Here we investigate the correlations between the X-ray and the multiwavelength properties of the sources in CAIXA. The physical implications of these results will be discussed within the framework of current AGN models, with particular interest on the open issues they still raise.

\section{\label{results}Results}

We have performed a systematic correlation analysis between all the parameters which characterise the sources in CAIXA. The results are summarized in Table \ref{corr}. In this paper, we consider statistically significant only the correlations whose Null Hypothesis Probability (NHP) is lower than $1\times10^{-3}$ (corresponding to a confidence level of 99.9\%), with respect to the Spearman correlation coefficient $\rho$ resulting from a censored fit (see Appendix \ref{appa} for details on this procedure, which takes into account both errors and upper limits on the Y variable).

\begin{table*}
\caption{\label{corr}Correlations between the main parameters which characterise the sources in CAIXA.}
\begin{center}
% use packages: array
\begin{tiny}
\begin{tabular}{c|cccccccccccc}
 & $L_\mathrm{2-10}$ & $L_\mathrm{0.5-2}$ & $L_\mathrm{0.5-2}/L_\mathrm{2-10}$ & EW$_{6.4}$ & $\Gamma_h$ & L$_\mathrm{bol}/L_\mathrm{Edd}$ & H$_\beta$ & M$_\mathrm{BH}$ & L$_\mathrm{6\,cm}$ & L$_\mathrm{20\,cm}$ & R$_\mathrm{x}$ & R$_\mathrm{l}$\\
\hline
\multirow{2}{*}{$L_\mathrm{2-10}$} & \multirow{2}{*}{--} & 0.92 & \multirow{2}{*}{X} & -0.33 & \multirow{2}{*}{X}& \multirow{2}{*}{X} & 0.37 & 0.80 & 0.65 & 0.79 & \multirow{2}{*}{X} & \multirow{2}{*}{X}\\
 & & $<-38$ &  & -4.4 & & & -4 & -20.1 & -11.7 & -23.0 & & \\
\multirow{2}{*}{$L_\mathrm{0.5-2}$} & 0.92 & \multirow{2}{*}{--} & \multirow{2}{*}{X} & -0.30 & \multirow{2}{*}{X} & \multirow{2}{*}{X} & \multirow{2}{*}{X} & 0.67 & 0.63 & 0.77 & \multirow{2}{*}{X} & \multirow{2}{*}{X}\\
& $<-38$ & & & -3.2 & & & & -10.7 & -10.7 & -23.1 & &\\
\multirow{2}{*}{$L_\mathrm{0.5-2}/L_\mathrm{2-10}$} & \multirow{2}{*}{X} & \multirow{2}{*}{X} & \multirow{2}{*}{--} & \multirow{2}{*}{X} & 0.42 & \multirow{2}{*}{X} & -0.64 & \multirow{2}{*}{X} & \multirow{2}{*}{X} & \multirow{2}{*}{X} & \multirow{2}{*}{X} & \multirow{2}{*}{X} \\
& & & & & -7.3 & & -11.2 & & & & & \\
\multirow{2}{*}{EW$_{6.4}$} & -0.33 & -0.30 & \multirow{2}{*}{X} & \multirow{2}{*}{--} & \multirow{2}{*}{X} & -0.38 & \multirow{2}{*}{X} & \multirow{2}{*}{X} & \multirow{2}{*}{X} & \multirow{2}{*}{X} & \multirow{2}{*}{X} & \multirow{2}{*}{X} \\
& -4.4 & -3.2 & & & & -3.2 & & & & & & \\
\multirow{2}{*}{$\Gamma_h$} & \multirow{2}{*}{X} & \multirow{2}{*}{X} & 0.42 & \multirow{2}{*}{X} & \multirow{2}{*}{--} & \multirow{2}{*}{X} & -0.39 & \multirow{2}{*}{X} & \multirow{2}{*}{X} & \multirow{2}{*}{X} & \multirow{2}{*}{X} & \multirow{2}{*}{X} \\ 
& & & -7.3 & & & & -4.1 & & & & & \\
\multirow{2}{*}{L$_\mathrm{bol}/L_\mathrm{Edd}$} & \multirow{2}{*}{X} & \multirow{2}{*}{X} & \multirow{2}{*}{X} & -0.38 & \multirow{2}{*}{X} & \multirow{2}{*}{--} & -0.40 & \multirow{2}{*}{X} & \multirow{2}{*}{X} & \multirow{2}{*}{X} & \multirow{2}{*}{X} & \multirow{2}{*}{X} \\ 
& & & & -3.2 & & & -3.7 & & & & & \\ 
\multirow{2}{*}{H$_\beta$} & 0.37 & \multirow{2}{*}{X} & -0.64 & \multirow{2}{*}{X} & -0.39 & -0.40 & \multirow{2}{*}{--} & 0.69 & \multirow{2}{*}{X} & \multirow{2}{*}{X} & \multirow{2}{*}{X} & \multirow{2}{*}{X}\\
& -4 & & -11.2 & & -4.1 & -3.7 & & -12.0 & & & &\\
\multirow{2}{*}{M$_\mathrm{BH}$} & 0.80 & 0.67 & \multirow{2}{*}{X} & \multirow{2}{*}{X} & \multirow{2}{*}{X} & \multirow{2}{*}{X} & 0.69 & \multirow{2}{*}{--} & 0.49 & 0.66 & \multirow{2}{*}{X} & \multirow{2}{*}{X}\\
 & -20.1 & -10.7 & & & & & -12.0 & & -4.7 & -8.0 & & \\
\multirow{2}{*}{L$_\mathrm{6\,cm}$} & 0.65 & 0.63 & \multirow{2}{*}{X} & \multirow{2}{*}{X} & \multirow{2}{*}{X} & \multirow{2}{*}{X}& \multirow{2}{*}{X} & 0.49 & \multirow{2}{*}{--} & 0.92 & 0.65 & 0.35 \\
& -11.7 & -10.7 & & & & & & -4.7 & & -25.3 & -11.4 & -3.2 \\
\multirow{2}{*}{L$_\mathrm{20\,cm}$} & 0.79 & 0.77 & \multirow{2}{*}{X} & \multirow{2}{*}{X} & \multirow{2}{*}{X} & \multirow{2}{*}{X} & \multirow{2}{*}{X} & 0.66 & 0.92 & \multirow{2}{*}{--} & 0.55 & \multirow{2}{*}{X}\\
& -23.0 & -23.1 & & & & & & -8.0 & -25.3 & & -9.5 & \\
\multirow{2}{*}{R$_\mathrm{x}$} & \multirow{2}{*}{X} & \multirow{2}{*}{X} & \multirow{2}{*}{X} & \multirow{2}{*}{X} & \multirow{2}{*}{X} & \multirow{2}{*}{X} & \multirow{2}{*}{X} & \multirow{2}{*}{X} & 0.65 & 0.55 & \multirow{2}{*}{--} & 0.58 \\
& & & & & & & & & -11.4 & -9.5 & & -14.0\\
\multirow{2}{*}{R$_\mathrm{l}$} & \multirow{2}{*}{X} & \multirow{2}{*}{X} & \multirow{2}{*}{X} & \multirow{2}{*}{X} & \multirow{2}{*}{X} & \multirow{2}{*}{X} & \multirow{2}{*}{X} & \multirow{2}{*}{X} & 0.35 & \multirow{2}{*}{X} & 0.58 & \multirow{2}{*}{--} \\
& & & & & & & & & -3.2 & & -14.0 & \\
\end{tabular}
\end{tiny}
\end{center}
For each couple, the Spearman correlation coefficient $\rho$ is reported (top), along with the logarithm of the NHP (bottom). If the latter is larger than -3.0, we consider the correlation with low significance and an `X' is shown. Parameters with no significant correlations, namely EW$_{6.7}$, EW$_{6.97}$ and $\Gamma_s$ are not listed in the table. Correlations involving $L_\mathrm{0.5-2}$ or $L_\mathrm{0.5-2}/L_\mathrm{2-10}$ were performed with the unobscured sub-catalogue of CAIXA. In this table, we also show trivial correlations for completeness, such as the $L_{0.5-2}/L_{2-10}$ versus $\Gamma_h$, or the H$\beta$ FWHM versus $L_{bol}/L_{Edd}$ one. See B09 and text for details.
\end{table*}

We will discuss in detail in the next sections all the significant correlations we found in CAIXA. 

\subsection{\label{hbeta}Correlations with H$\beta$ FWHM and the nature of the soft excess}

\begin{figure}
\begin{center}
\epsfig{file=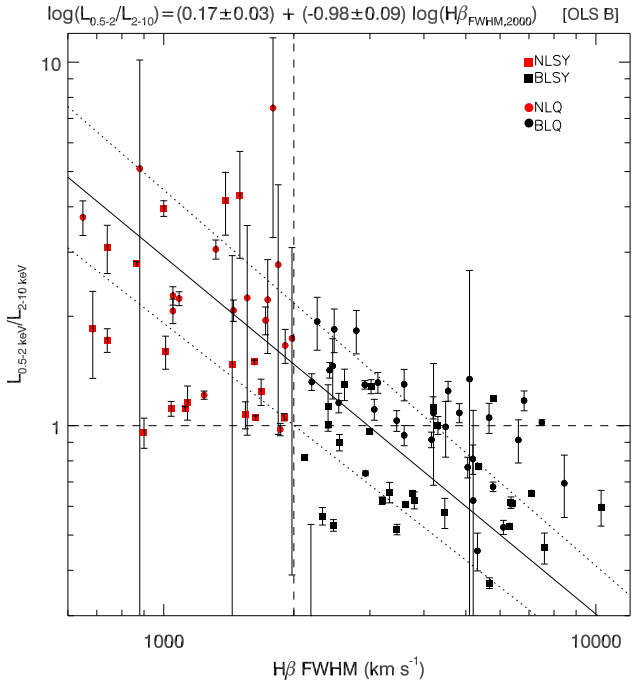, width=9.1cm}
\end{center}
\caption{\label{lxratio_hbeta}\textit{Left}: 0.5-2/2-10 keV luminosity ratio vs. H$\beta$ FWHM. The anti-correlation between the two parameters is very significant. The two broken lines represent the 2000 km s$^{-1}$ boundary between broad- and narrow-line objects and $L_{0.5-2}/L_{2-10}=1$. See caption of Fig. \ref{lx_lradio} for details on the adopted symbols.}
\end{figure}

Looking at Table \ref{corr}, the most significant correlation with the FWHM of H$\beta$ is that against the ratio between the soft (0.5-2 keV) and the hard (2-10 keV) X-ray luminosity. The observed anti-correlation is shown in Fig. \ref{lxratio_hbeta}.

Given the unknown origin of the soft excess in AGN, we introduced in B09 this X-ray luminosity ratio as a good parameter to characterize at least the basic property of the soft excess of the sources, i.e. its strength with respect to the primary X-ray emission. As already noted in B09, it is very interesting to note that CAIXA does not contain narrow-line objects with a $L_{0.5-2}/L_{2-10}$ ratio lower than 1 and narrow- and broad-line objects are significantly different populations with respect to this parameter. However, just as no dichotomy appears in the optical properties between narrow- and broad-line objects, a smooth transition between high and low X-ray luminosity ratios is also apparent from the data. The best fit for the anti-correlation takes this form:

\begin{equation}
\log(\frac{L_{0.5-2}}{L_{2-10}}) = \left(0.17\pm0.03\right) + \left(-0.98\pm0.09\right) \log(\mathrm{H\beta_{2000}})
\end{equation}

\noindent where $\mathrm{H\beta_{2000}}$ is the FWHM expressed in units of $2\,000$ km s$^{-1}$. We remind the reader that this correlation, which involves the soft X-ray luminosity, was calculated on the CAIXA sub-catalogue which excludes objects with a measured column density in excess of the Galactic one (see B09).

The other correlation with H$\beta$ found in CAIXA is the one with the hard X-ray spectral slope, confirming previous claims \citep[e.g.][]{bme97,pico05}. An anti-correlation between the 0.1-2.4 keV slope and the H$\beta$ line width was also presented by \citet{laor94} and later confirmed by \citet{bbf96} and \citet{laor97}. However, we note here that in CAIXA there is no significant anti-correlation between the H$\beta$ and the soft X-ray photon index. This is likely another piece of evidence in favour of the fact that the values of $\Gamma$ are very sensitive to the adopted models and it is therefore better to consider the correlations between more model-independent parameters. We therefore believe that our anti-correlation between the X-ray luminosity ratio and H$\beta$ is a more basic physical correlation, because it links two \textit{a priori} independent properties of radio-quiet AGN.

The transition from broad- to narrow-line AGN, as well as the strength of the soft excess have been often interpreted as being due to a more fundamental physical parameter, such as the accretion rate or the BH mass \citep[e.g.][]{laor97}. However, there is no significant correlation in CAIXA between $L_{0.5-2}/L_{2-10}$ and the Eddington ratio, the BH mass or any parameter considered in this work other than the FWHM of H$\beta$ (a part from the one with $\Gamma_\mathrm{h}$ discussed above). In other words, we do not have any indications that the strength of the soft excess may be related to any other important parameter, a part from the FWHM of H$\beta$, not even the accretion rate.

The `classical' interpretation of the soft excess as direct emission from the accretion disc was questioned when several studies \citep[e.g.][]{gd04,crummy06} showed that the observed temperature of resulting black body is remarkably constant across orders of magnitude of luminosities and BH masses. We confirmed the existence of a `universal' temperature, independent on luminosity or BH mass, on the large number of sources of CAIXA, if we substitute in the baseline model the soft X-ray power-law with a thermal component (B09). In the last few years, two main competing models have been proposed to explain the origin of the soft excess, both ascribing its apparent universal temperature to atomic physics processes. In one scenario, the soft excess arises from the enhancement of reflection from the inner regions of the accretion disc due to light-bending effects, together with a strong suppression of the primary emission \citep[e.g.][]{mf04}. On the other hand, the soft excess can also be mimicked by absorption from a relativistically outflowing warm gas \citep[e.g.][]{gd04}. Regardless of the details of the two models \citep[and see also][for a self-consistent model of X-ray emission from inhomogeneous, clumpy discs]{merl06b}, they both imply that the observed X-ray emission is suppressed with respect to the intrinsic one. 

\begin{figure*}
\begin{center}
\epsfig{file=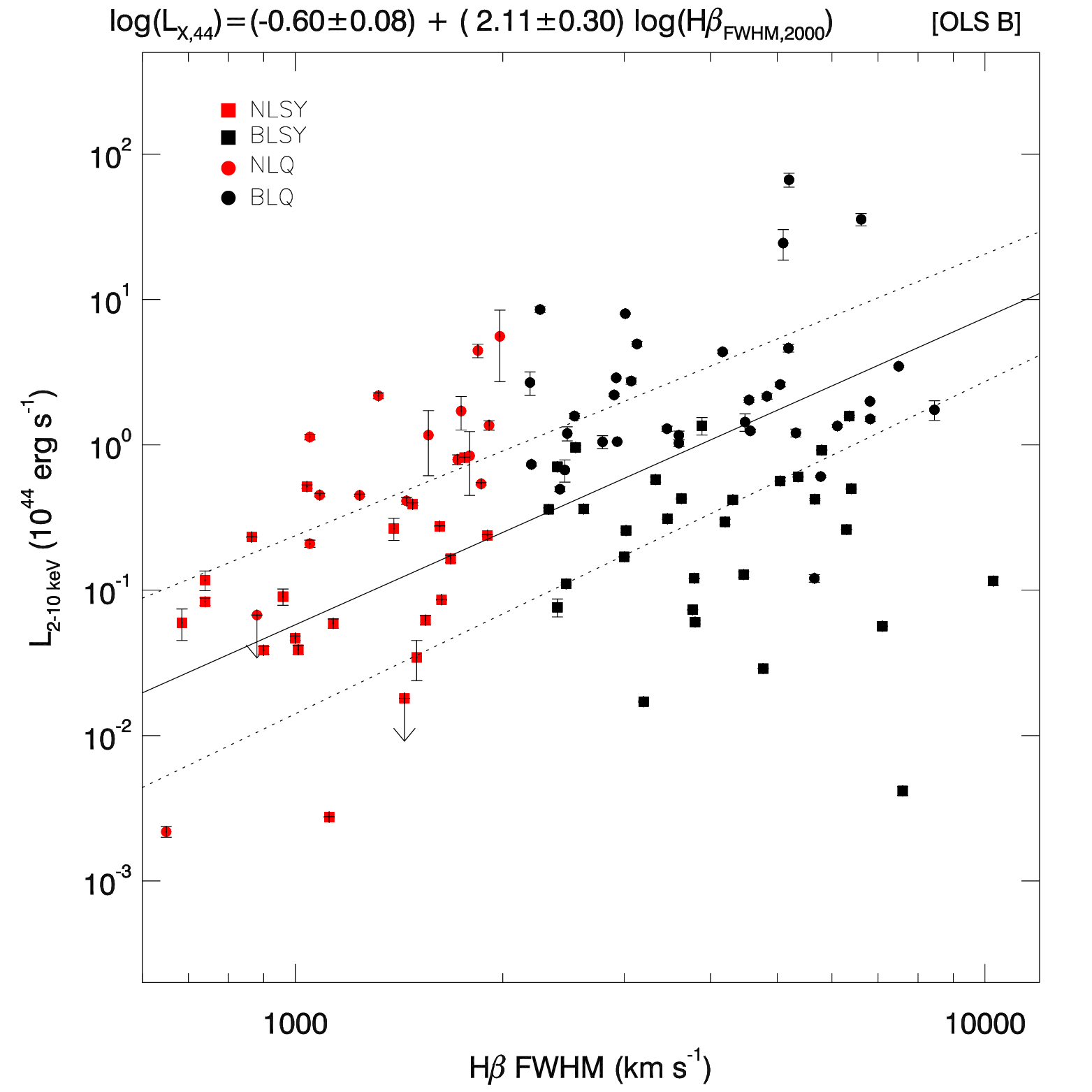, width=9.1cm}
\epsfig{file=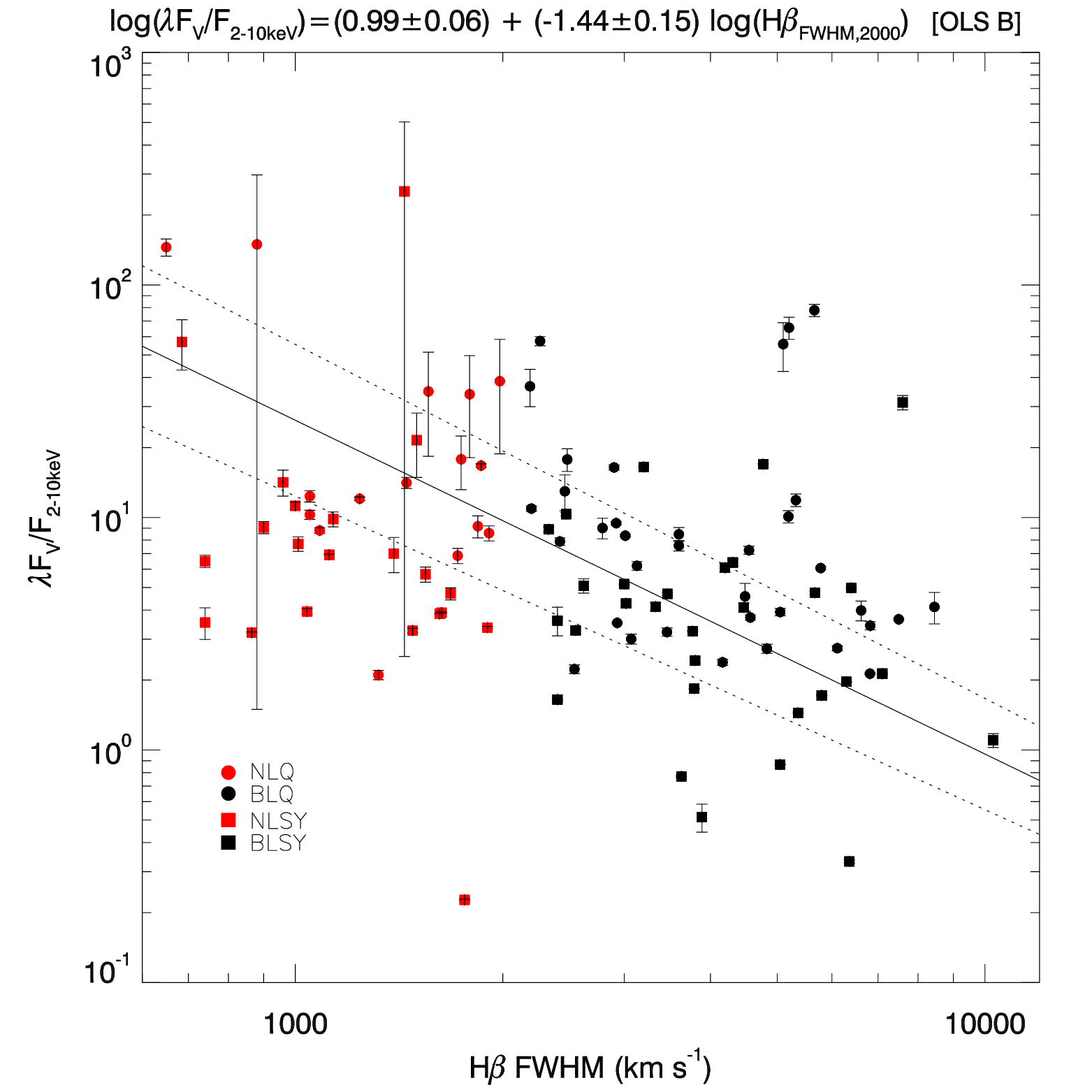, width=9.1cm}
\end{center}
\caption{\label{vxratio_hbeta}\textit{Left panel}: Hard X-ray luminosity vs FWHM of H$\beta$: the two parameters appear correlated. \textit{Right panel}: Ratio between the optical flux (V filter) and the hard X-ray flux against the FWHM of H$\beta$: the observed anti-correlation suggests that narrow-line objects may be X-ray under-luminous. See text for details and caption of Fig. \ref{lx_lradio} for details on the adopted symbols.}
\end{figure*}

Therefore, the two models predict that the soft excess should be more prominent in X-ray weaker sources. As we have already mentioned, there is no significant correlation between the X-ray luminosity ratio (our proxy for the strength of the soft excess) and any other parameter, apart from the FWHM of the H$\beta$. We can only look for indications that the latter is correlated to any observed suppression of the intrinsic X-ray continuum. Indeed, we do find a significant correlation between the hard X-ray luminosity and the FWHM of H$\beta$ (see Table \ref{corr} and Fig. \ref{vxratio_hbeta}), confirming previous claims \citep[e.g.][]{pico05}. And, more interestingly, we also find that the ratio between the optical flux in the V band and the hard X-ray flux is significantly anti-correlated ($\rho=-0.36$ - NHP=$2\times10^{-4}$) with the width of the H$\beta$. This clearly shows that the correlation of  $L_{0.5-2}/L_{2-10}$ ratio with the H$\beta$ FWHM is driven by the hard X-ray luminosity getting weaker, rather than the soft X-ray luminosity getting stronger. Therefore, there are observational indications that objects with narrower H$\beta$ are X-ray weaker than normal broad-line sources. However, it is impossible from these results to disentangle between an absorption or reflection origin of the soft excess. Only detailed modeling of broadband spectra (including high SNR data above 10 keV) can solve the issue \citep[e.g.][]{ponti08}.

\subsection{\label{bhcorr}Correlations with BH mass and the X-ray bolometric correction}

\begin{figure*}
\begin{center}
\epsfig{file=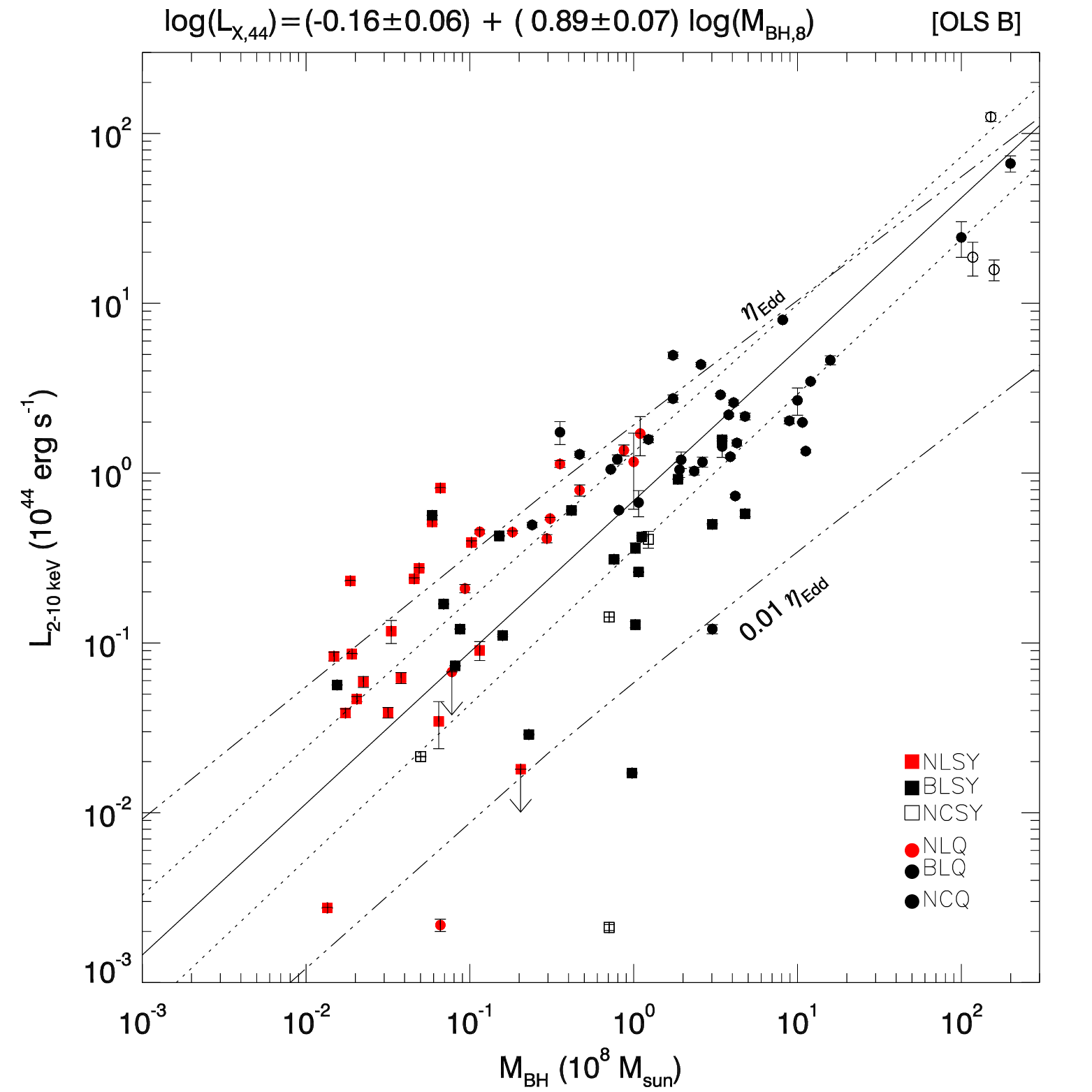, width=9.1cm}
\epsfig{file=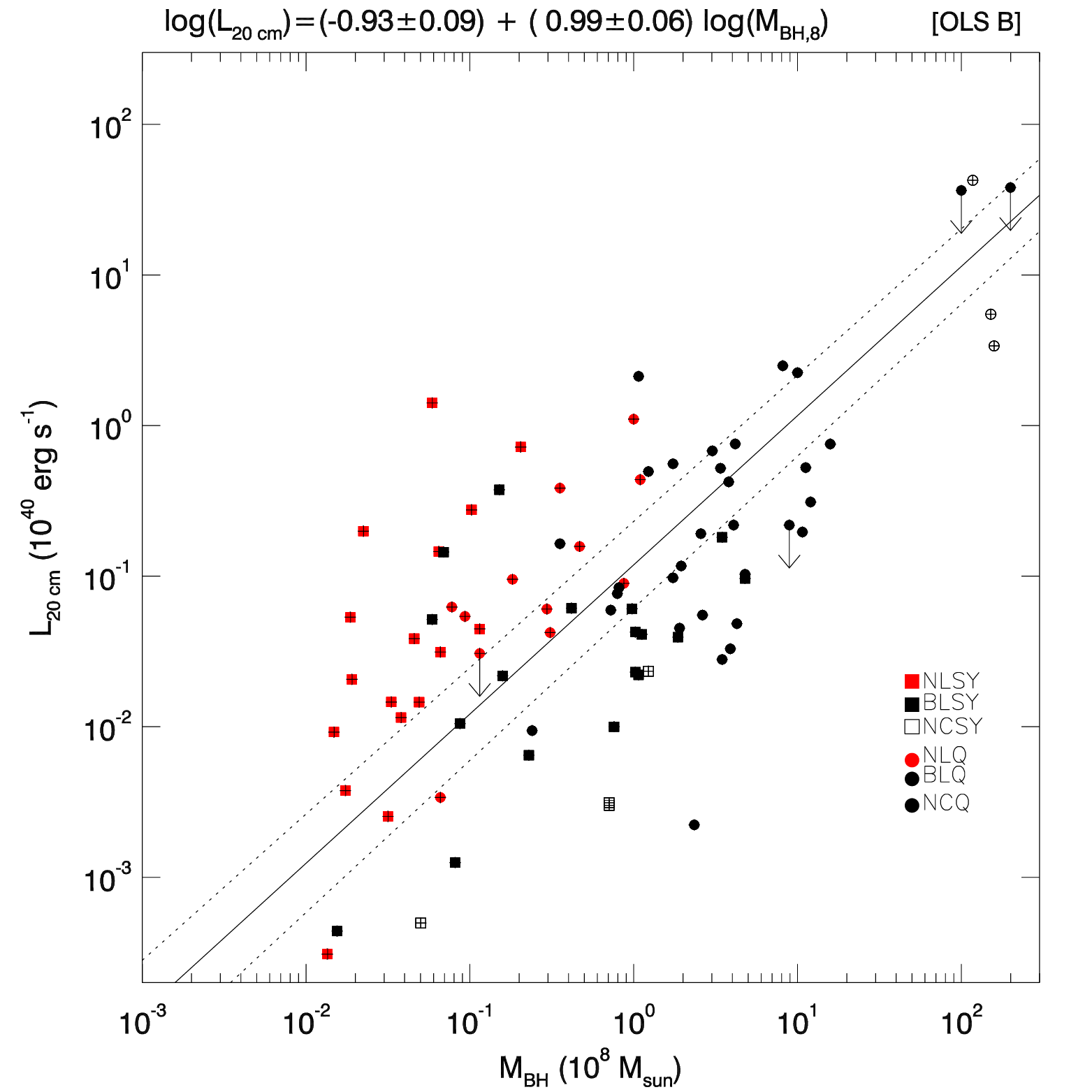, width=9.1cm}
\epsfig{file=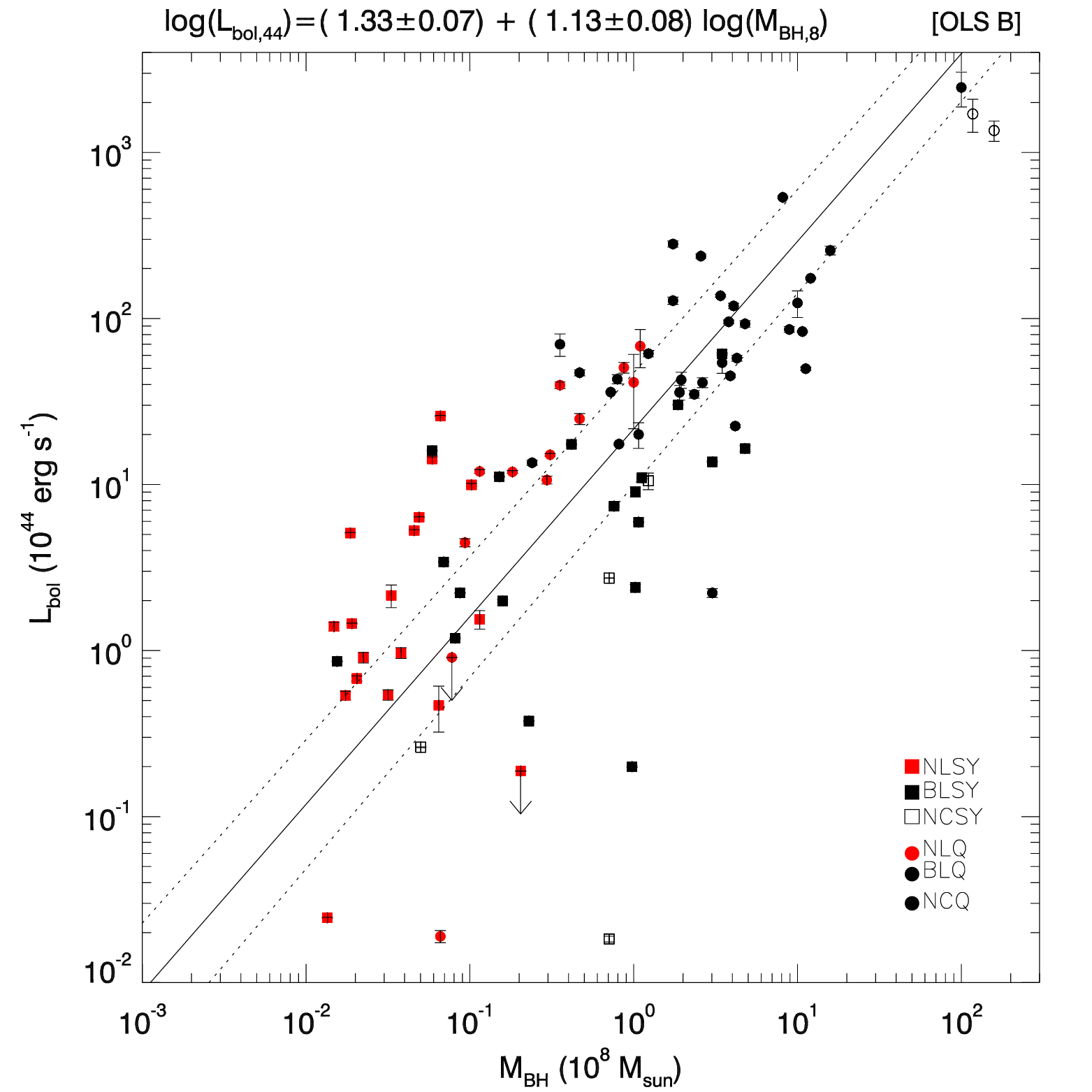, width=9.1cm}
\epsfig{file=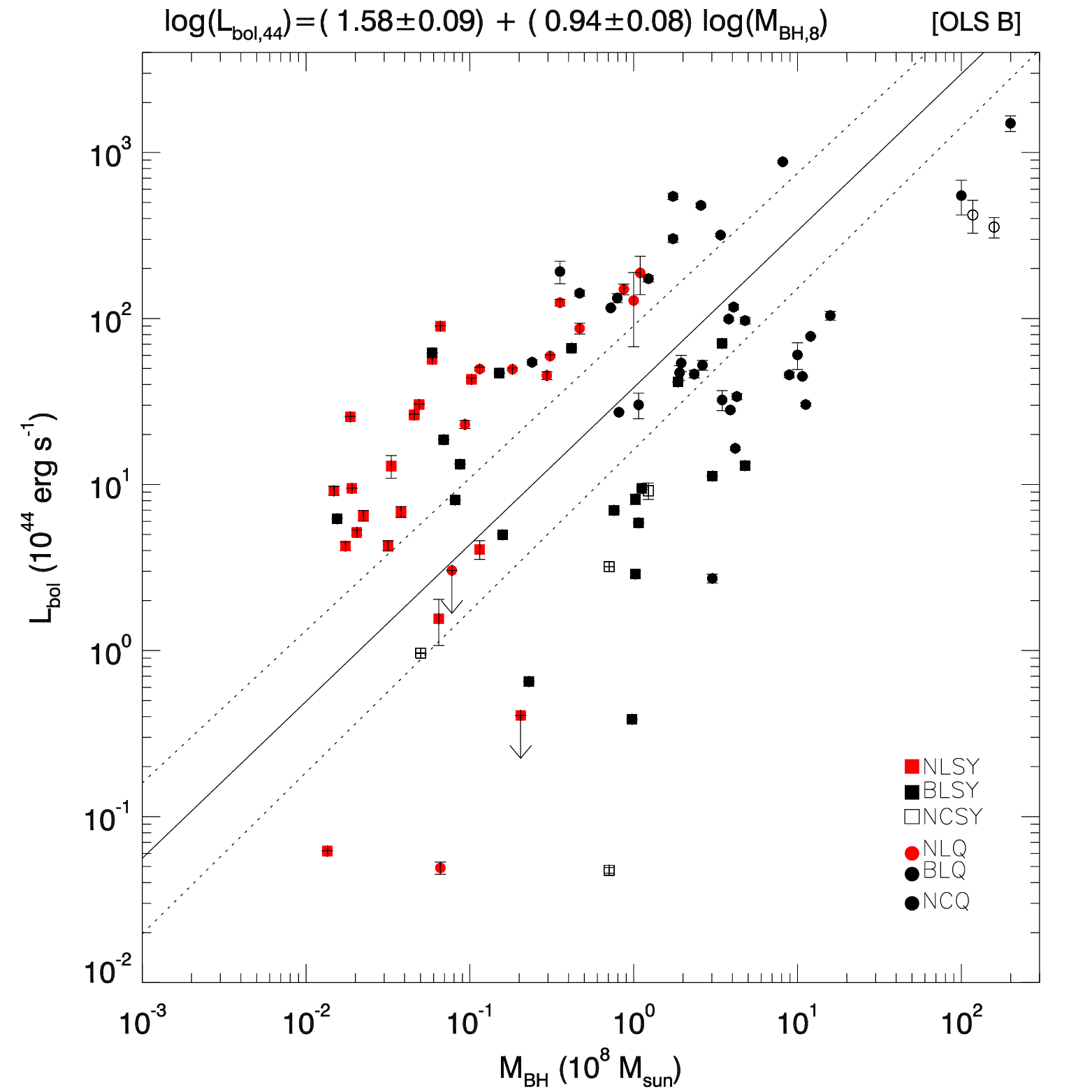, width=9.1cm}
\end{center}
\caption{\label{lxmbh}\textit{Top} 2-10 keV X-ray and 20 cm radio luminosity vs. BH mass for all the sources in the catalogue with an estimate of the latter. A linear regression fit is superimposed on the plots. In the left panel, the curves referring to Eddington rates of 1 and 0.01 are also shown. \textit{Bottom} Bolometric luminosity (\textit{left}: luminosity-dependent \citep{mar04}; \textit{right}: Eddington ratio-dependent \citep{vf09}) vs. BH mass for all the sources in the catalogue with an estimate of the latter. A linear regression fit is superimposed on the plots. See text for details.}
\end{figure*}

The strongest correlation with the BH mass in CAIXA is the one with the hard X-ray luminosity (see Table \ref{corr} and Fig. \ref{lxmbh}). This correlation is expected, because, if the accretion rate is a universal constant, the bolometric luminosity should rise linearly with the BH mass. Assuming that the bolometric correction, i.e. the ratio between the bolometric and the X-ray luminosity, is constant \citep[as in ][]{elvis94}, the correlation should be also linear with X-ray luminosity. The best fit slope for our data is slightly lower than 1, being $0.89\pm0.07$, in agreement with previous results \citep{ww09}. This flatter slope could suggest that objects with larger BH masses have intrinsically lower accretion rates or are simply X-ray weaker with respect to their bolometric luminosity. On the other hand, the correlation with the radio luminosity ($\rho=0.59$ - NHP=$2\times10^{-9}$ for the 20 cm catalogue) is consistent with a linear relation with the BH mass (see Fig. \ref{lxmbh}), thus supporting the hypothesis that the total power of the AGN scales linearly with the BH mass, while high-luminosity objects may need a larger X-ray bolometric correction factor.

The increasing optical-to-X-ray ratio with increasing optical luminosity is a well established result in AGN research \citep[see e.g.][and references therein]{at86,just07}. This is in agreement with the X-ray bolometric correction factor increasing with luminosity adopted in CAIXA, after \citet{mar04}. However, the correlation between the BH mass and the bolometric luminosity calculated in this way ($\rho=0.80$ - NHP=$7\times10^{-21}$) gives a slope steeper than 1, suggesting that the luminosity dependence of this correction factor may be too strong at large luminosities. Recently, \citet{vf09} have suggested that the X-ray bolometric correction may be dependent on the Eddington ratio, not the luminosity. We tried to adopt the average bolometric correction reported in their paper for each Eddington ratio bin. The result is shown in the bottom right plot of Fig. \ref{lxmbh}: a correlation between the bolometric corrected X-ray luminosity and the BH mass with a slope consistent with 1 is recovered.

\begin{figure}
\begin{center}
\epsfig{file=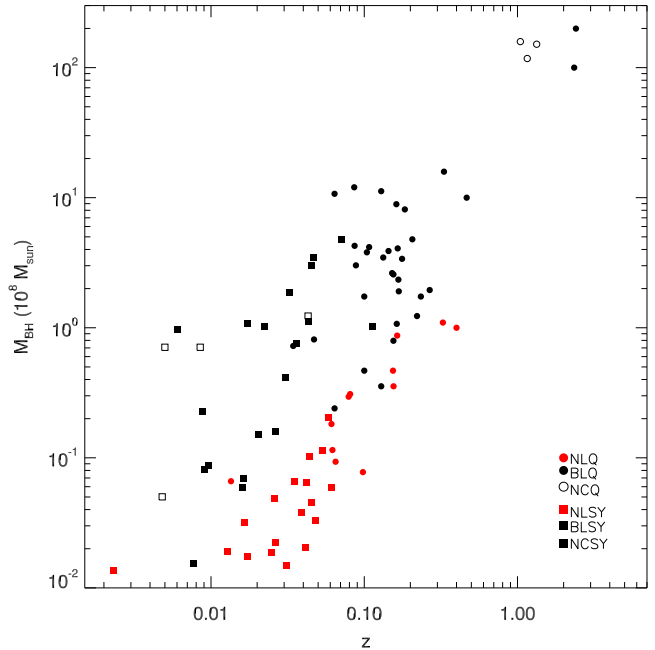, width=9.1cm}
\end{center}
\caption{\label{bh_bias}BH mass vs redshift. See text for details and caption of Fig. \ref{lx_lradio} for details on the adopted symbols.}
\end{figure}

Two other significant correlations with BH mass are found in CAIXA. The first is the correlation between the BH mass and z (see Fig. \ref{bh_bias}). This is expected: if the efficiency of the conversion of gravitational energy into radiative output is the same for all AGN, the objects at high z, to be in CAIXA, must be luminous and, likely, with a large BH mass. Note that, at each redshift, narrow-line objects have invariably the smallest BH masses: this is well known and some authors believe that the masses in these objects are underestimated \citep[e.g.][]{decarli08,mar08}. This effect represents the other correlation with the BH mass found in CAIXA and can be explained in terms of the dependence of the BH mass on the H$\beta$ FWHM. This is of course due to the fact that most of the BH mass in CAIXA were primarily estimated from the H$\beta$ FWHM under the assumption of a virialised Broad Line Region (BLR). This would imply $\mathrm{M_{BH}\propto FWHM(H\beta)^2}$. The fact that the actual correlation in CAIXA requires a slightly steeper slope ($2.96\pm0.23$) is likely the result of the inhomogeneous provenience of H$\beta$ data and methods for estimation of the BH masses.

No other significant correlations with BH mass are found in CAIXA. In particular, the correlation with the Eddington ratio is lower than our threshold, even if larger than 99\% confidence level ($\rho=-0.29$ - NHP=$6\times10^{-3}$).

\subsection{\label{radiocorr}Correlations with radio emission and its origin in radio-quiet AGN}

In this section, we discuss the correlations between the radio properties of the objects in CAIXA and any other parameter. It should be borne in mind that the catalogue does not include radio-loud objects, as explained in detail in B09.

\begin{figure*}
\begin{center}
\epsfig{file=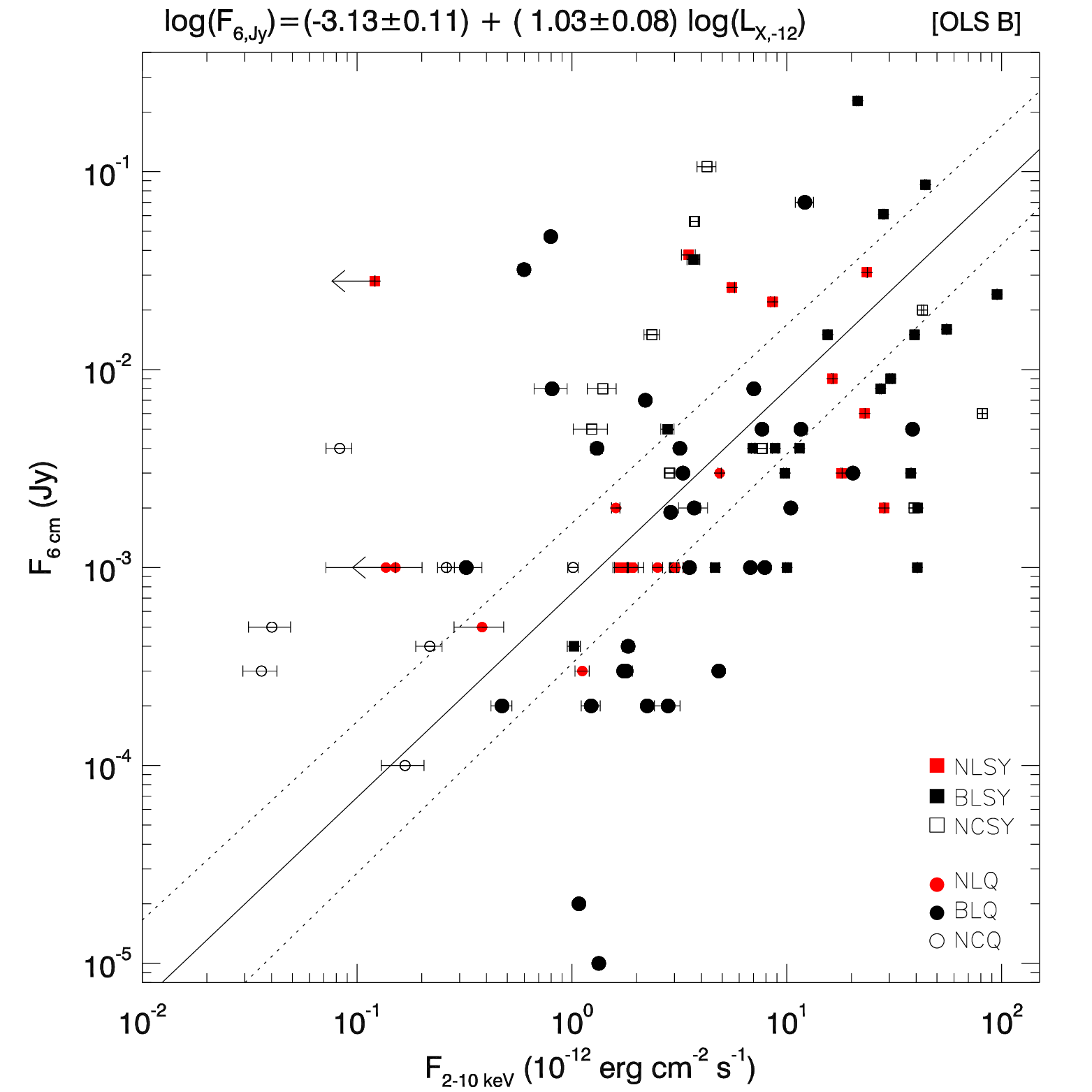, width=9.1cm}
\epsfig{file=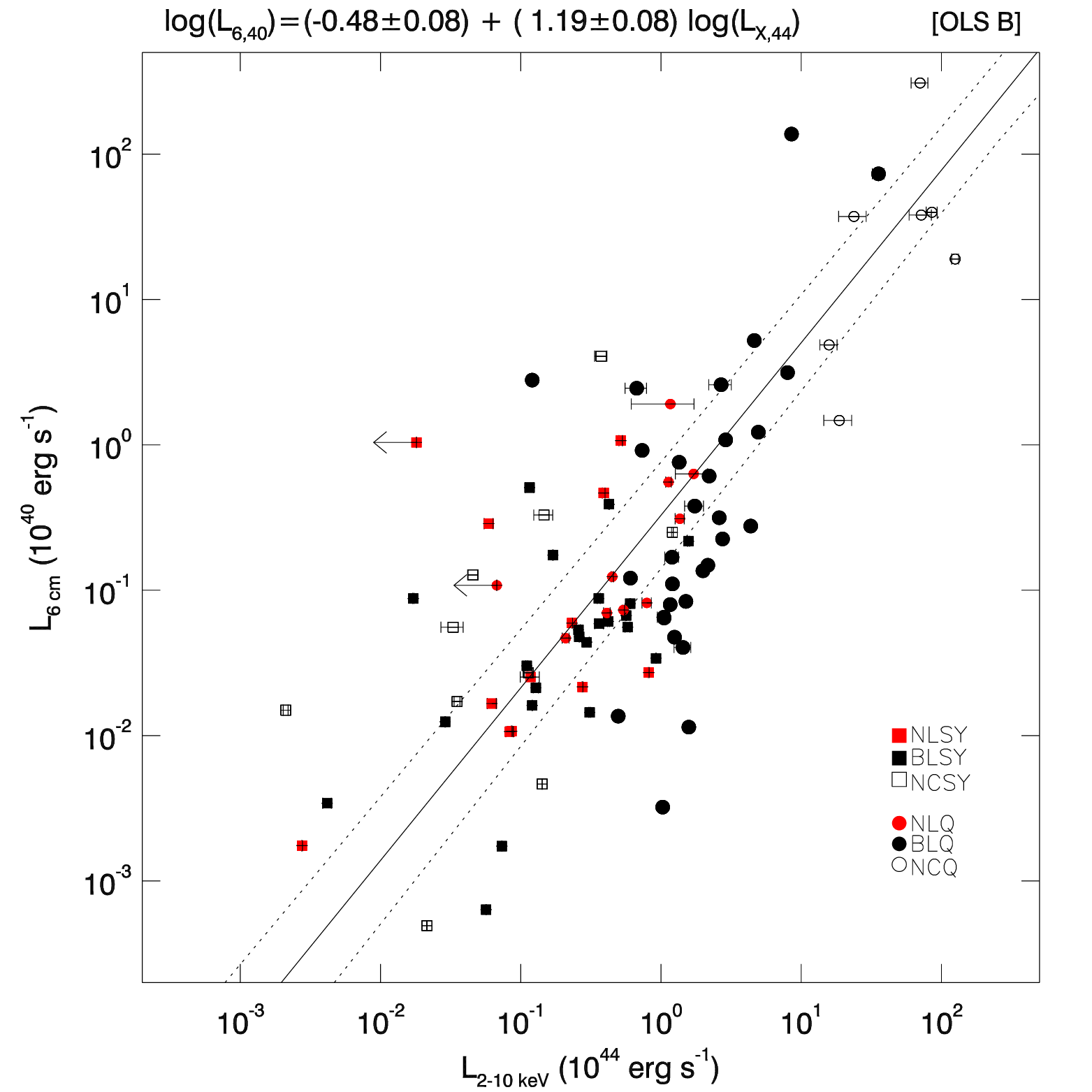, width=9.1cm}
\epsfig{file=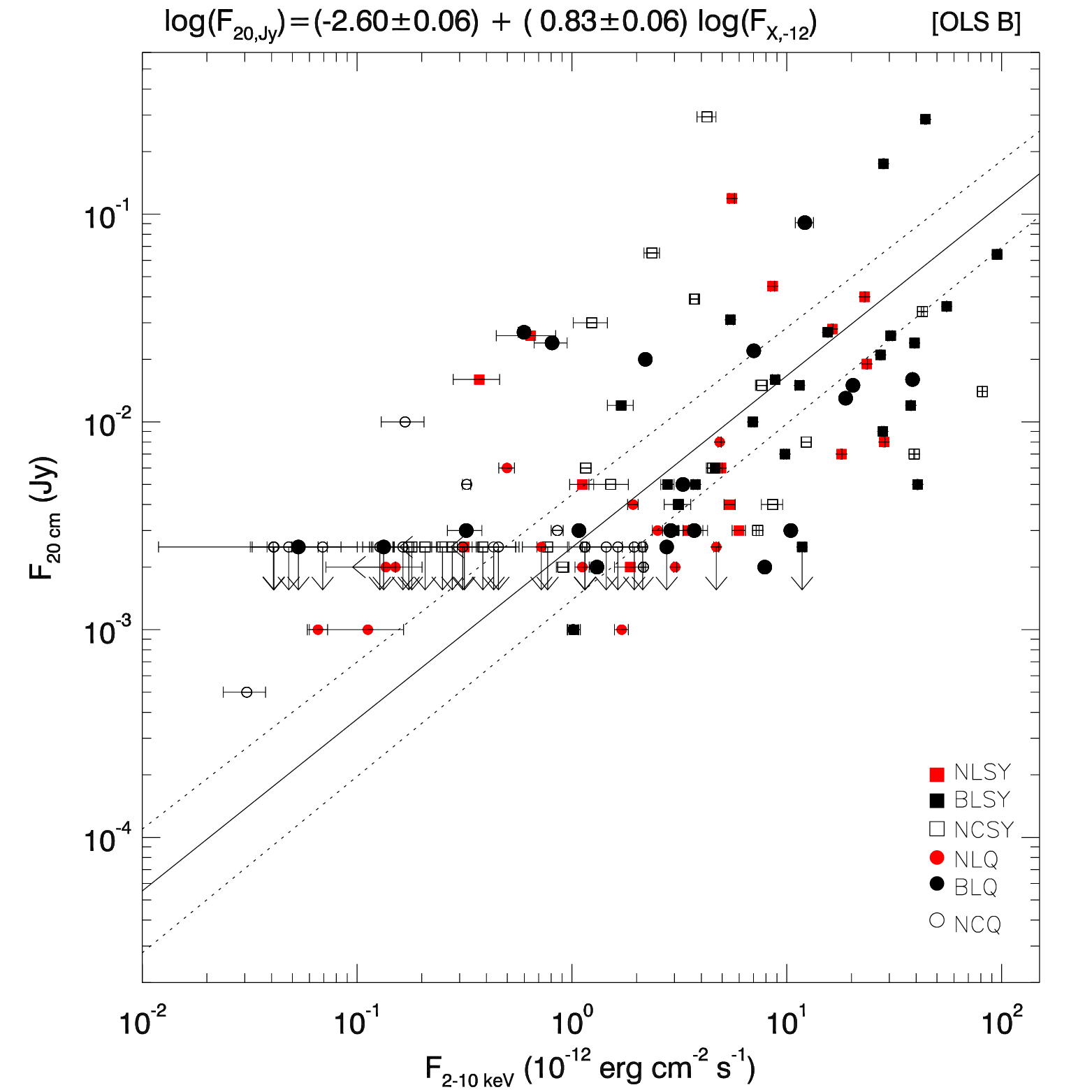, width=9.1cm}
\epsfig{file=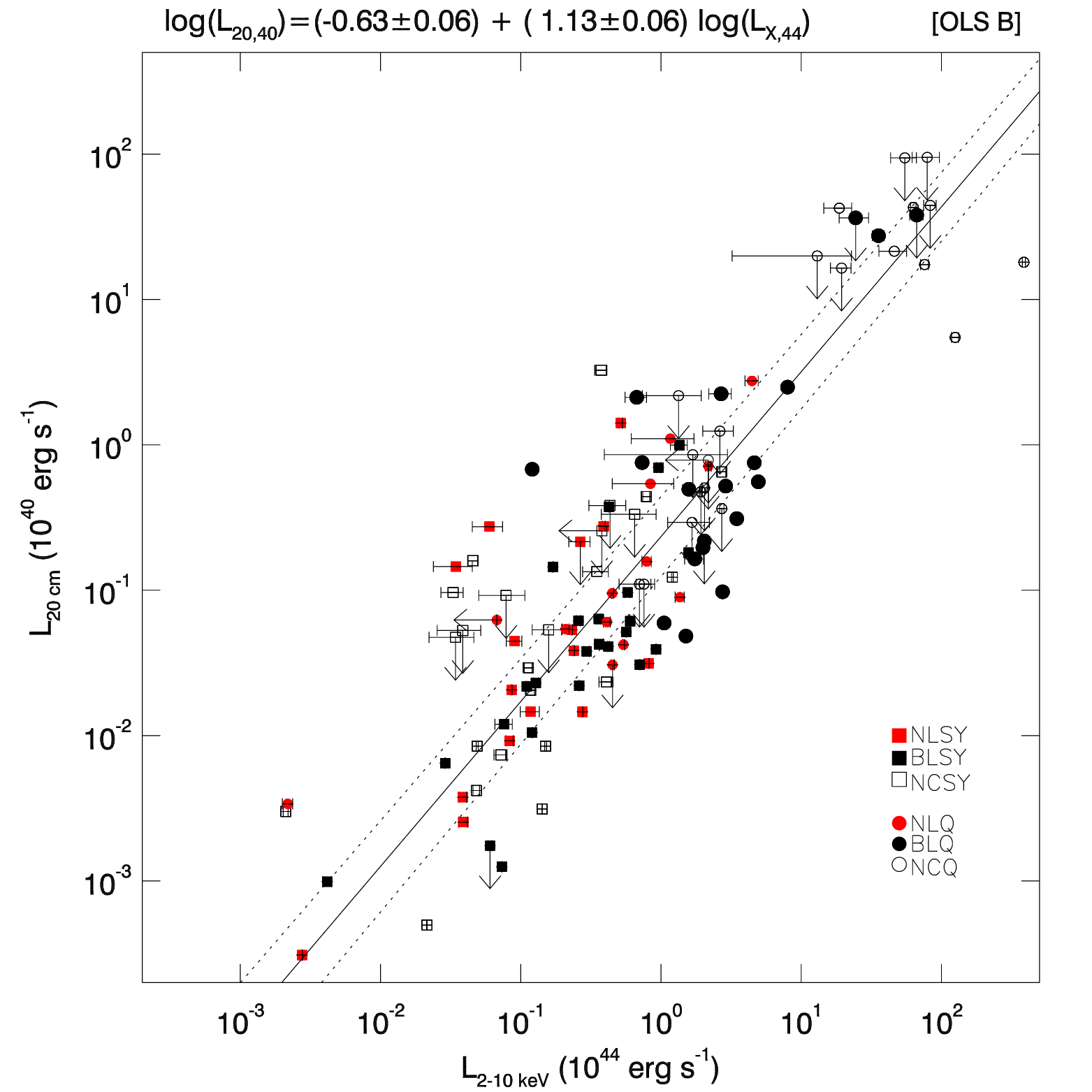, width=9.1cm}
\end{center}
\caption{\label{lx_lradio}Radio to X-rays plots. From top to bottom: 2-10 keV vs 6 cm emission, 2-10 keV vs 20 cm emission. Flux-flux correlations are on the left, luminosity-luminosity correlations on the right. In each case, the analytical expression for the best fit is on the top. The different symbols refers to the classification of the objects, on the basis of their absolute magnitude and H$\beta$ FWHM: \textit{NLSY}, narrow-line Seyfert 1; \textit{BLSY}, broad-line Seyfert 1; \textit{NCSY}, not-classified Seyfert 1 (no H$\beta$ FWHM measure available); \textit{NLQ}, narrow-line quasar; \textit{BLQ}, broad-line quasar; \textit{NCQ}, not-classified quasar (no H$\beta$ FWHM measure available).}
\end{figure*}

In Fig. \ref{lx_lradio} (right panels), we plot the hard X-ray versus the radio luminosities (at 6 and 20 cm) for the objects in our catalogue. The correlation is highly significant at both wavelengths (see Table \ref{corr}), with the following best fits:

\begin{equation}
\log(L_{6,40}) = \left(-0.48\pm0.08\right) + \left(1.19\pm0.08\right) \log(L_{2-10,44})
\end{equation}
\begin{equation}
\log(L_{20,40}) = \left(-0.63\pm0.06\right) + \left(1.13\pm0.06\right) \log(L_{2-10,44})\\
\end{equation}

\noindent where radio and X-ray luminosities are given in units of $10^{40}$ and $10^{44}$ erg s$^{-1}$, respectively.

While the radio emission of radio-loud objects has been clearly established as originating in their relativistic jets \citep[e.g.][]{bbr84}, the radio activity in radio-quiet sources is still puzzling. Several authors have shown that X-ray and radio luminosity strongly correlate in radio-quiet AGN \citep[e.g.][]{brink00,panessa07,lww08,lb08}. This result is generally interpreted as evidence that also in radio-quiet objects radio emission can be considered a direct tracer of the nuclear activity, either through a common accretion process or through the scale-invariant dependence of the radio-emitting jet on the accretion physics \citep[e.g.][]{hs03,merl03}.

\begin{figure}
\begin{center}
\epsfig{file=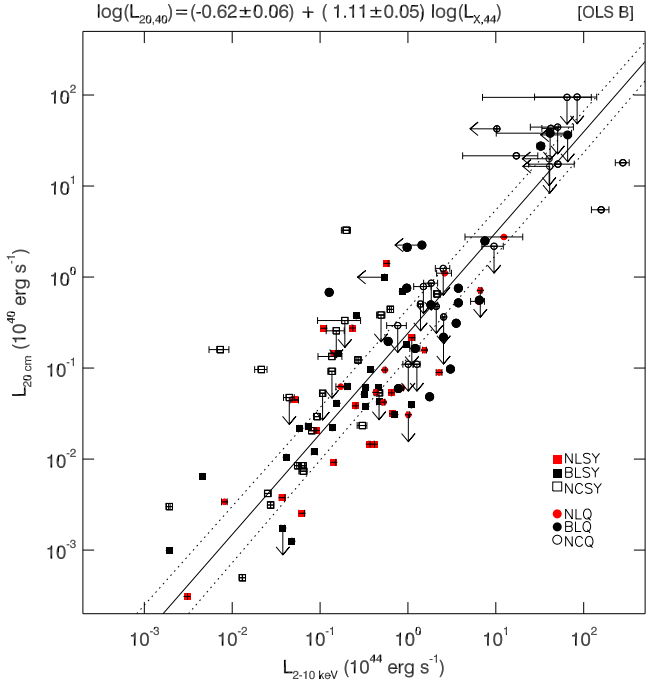, width=9.1cm}
\end{center}
\caption{\label{lxsoft_lradio}Radio 20 cm vs. soft X-ray luminosity, along with best fit. See caption of Fig. \ref{lx_lradio} for details on the adopted symbols.}
\end{figure}

The presence of the strong correlation found in CAIXA is in agreement with previous results, even if the slope of our best fits (around 1.1-1.2) is somewhat steeper than the linear relations found by \citet{brink00}, \citet{panessa07} and \citet{lb08} for radio-quiet objects. Our slope remains steeper even if we consider the correlation between the radio emission and the soft X-ray luminosity (see Table \ref{corr} and Fig. \ref{lxsoft_lradio}), as the above mentioned authors do \citep[with the exception of][]{panessa07}. Other authors found correlations far from being linear, and significantly different between radio-loud and radio-quiet objects \citep[e.g.][]{lww08}. We therefore decided to investigate possible biases affecting this correlation.

It is known that care must be taken in assessing the real significance of any luminosity-luminosity correlation and in giving a physical meaning to the best fit slope, because of the effects introduced by distance and the selection of the catalogue. Indeed, as expected, the luminosity is tightly correlated to the redshift, due to the sensitivity limit of our observations (although it varies depending on the exposure time of each observation).

In the left panels of Fig. \ref{lx_lradio}, we show the same correlations as above, but with fluxes instead of luminosity. The correlations are still statistically significant, but the significance is lower ($\rho=0.49$ and 0.67, with NHP=$7\times10^{-7}$ and $5\times10^{-16}$, for $F_6$ and $F_{20}$, respectively) and the scatter is much larger than that between luminosities. Moreover, the best slopes are flatter than the ones found with luminosities, in particular for the measures at 20 cm, where there are many upper limits. This suggests that the role of distance in the previous luminosity-luminosity plots is indeed important. The reason is obvious: only objects with high luminosities in both bands can be seen at large distances, while both high-and-low luminosity objects can be seen at small distances (but the small volume often reduces the number of high-L objects). This was shown in the simulations by \citet{isobe86}, who demonstrated that, if the upper limits are kept, then survival analysis methods can in principle recover the correct relationship. However, our catalogue can still be affected by this effect, because, even if we keep upper limits (both on radio and X-ray fluxes), neither the X-ray nor the radio data-sets are by any means complete and they are selected only on the basis of the SNR of the first, thus likely introducing a bias towards high flux sources. We therefore do not consider the existence of a significant correlation between fluxes a proof significant enough that the distance bias does not play any role in the luminosity-luminosity correlation in CAIXA.

Accordingly, we performed a partial Kendall $\tau$ correlation test on censored data, using the distance as the third variable, as proposed by \citet{as96}. Also taking into account the correlation with distance, the X-ray versus radio luminosity correlation still remains significant in all the cases: coefficient $\tau=0.24$ and variance $\sigma=0.056$ (NHP=$3\times10^{-5}$) for 6 cm, $\tau=0.22$ and $\sigma=0.046$ (NHP=$1\times10^{-6}$) for 20 cm. 

\begin{figure*}
\begin{center}
\epsfig{file=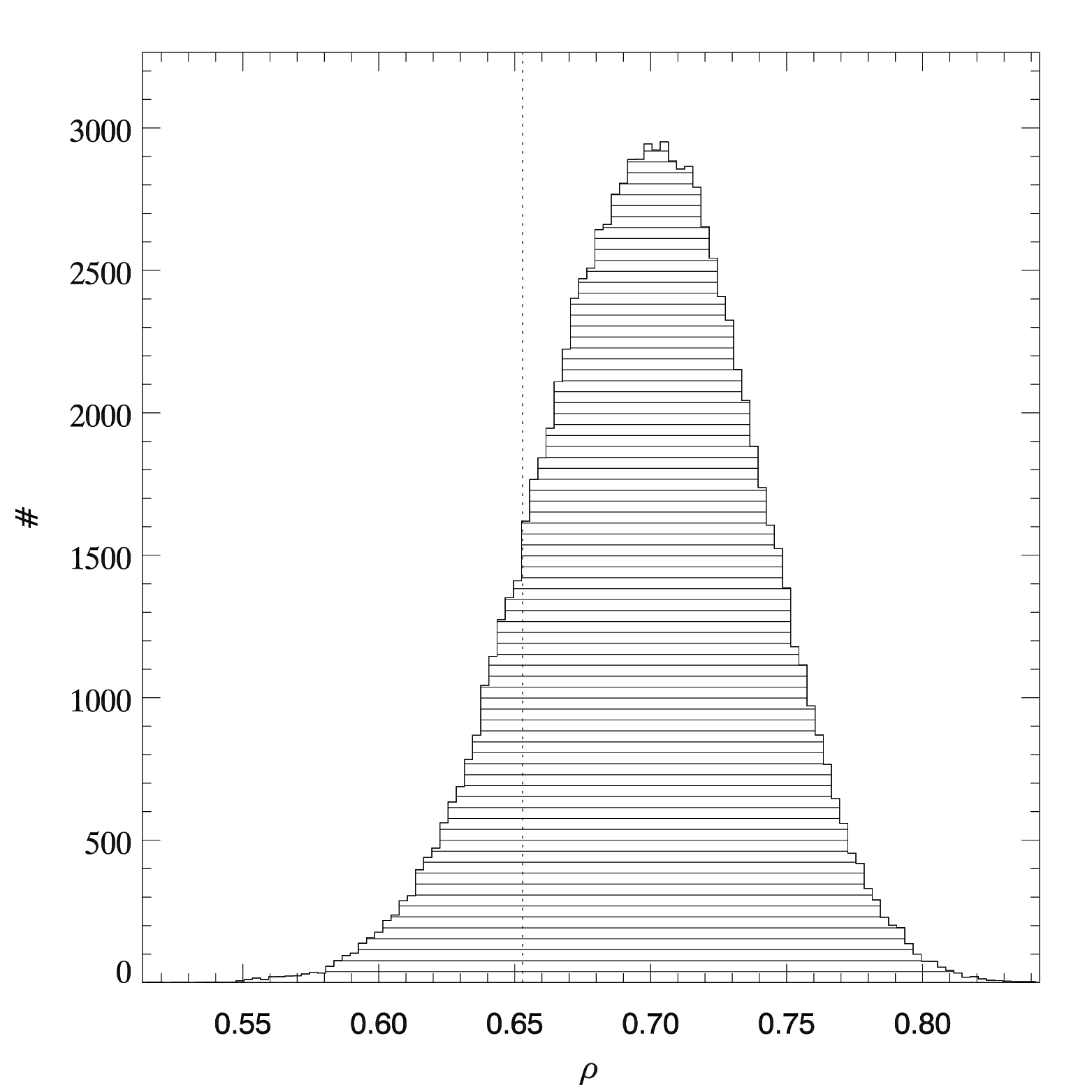, width=9.1cm}
\epsfig{file=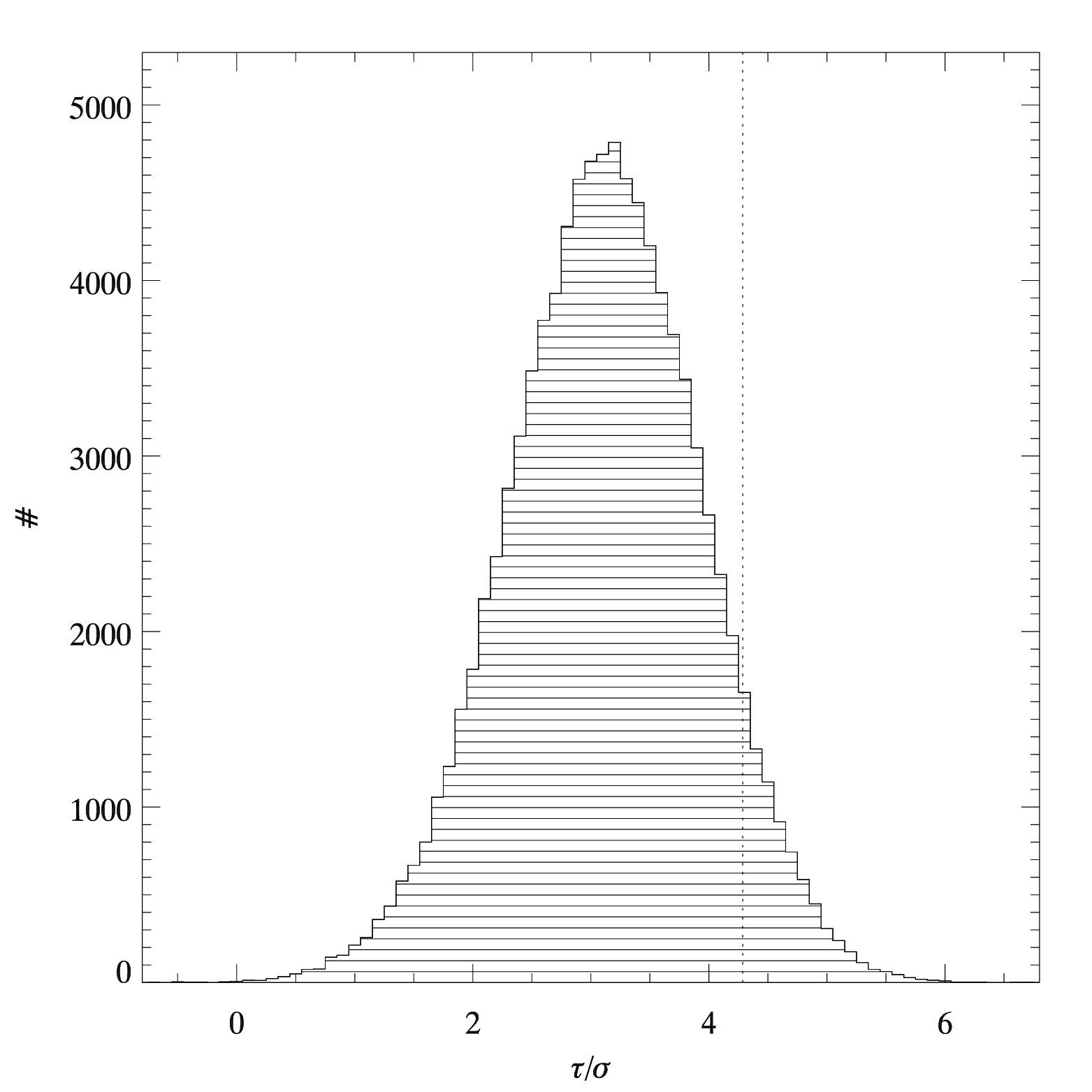, width=9.1cm}
\epsfig{file=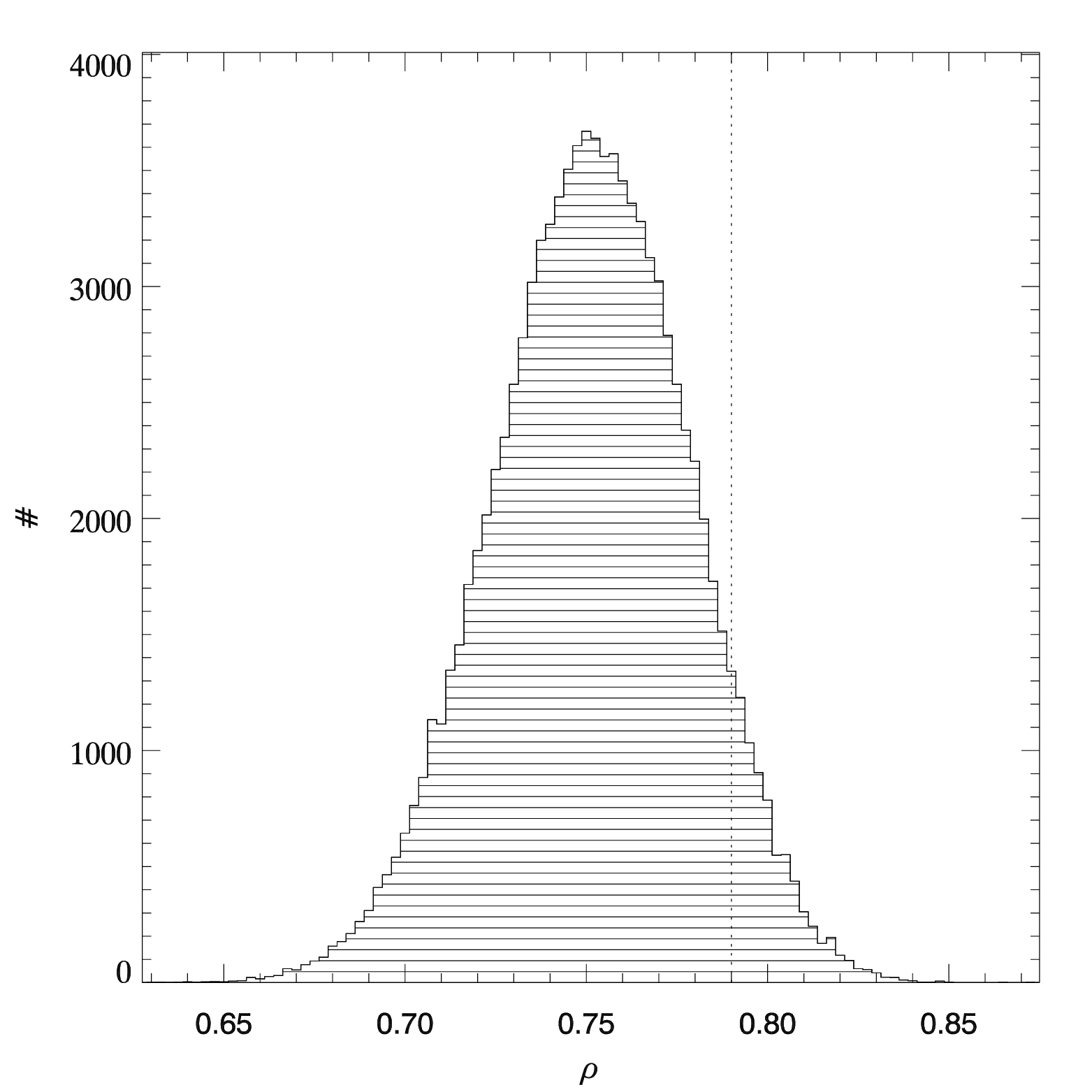, width=9.1cm}
\epsfig{file=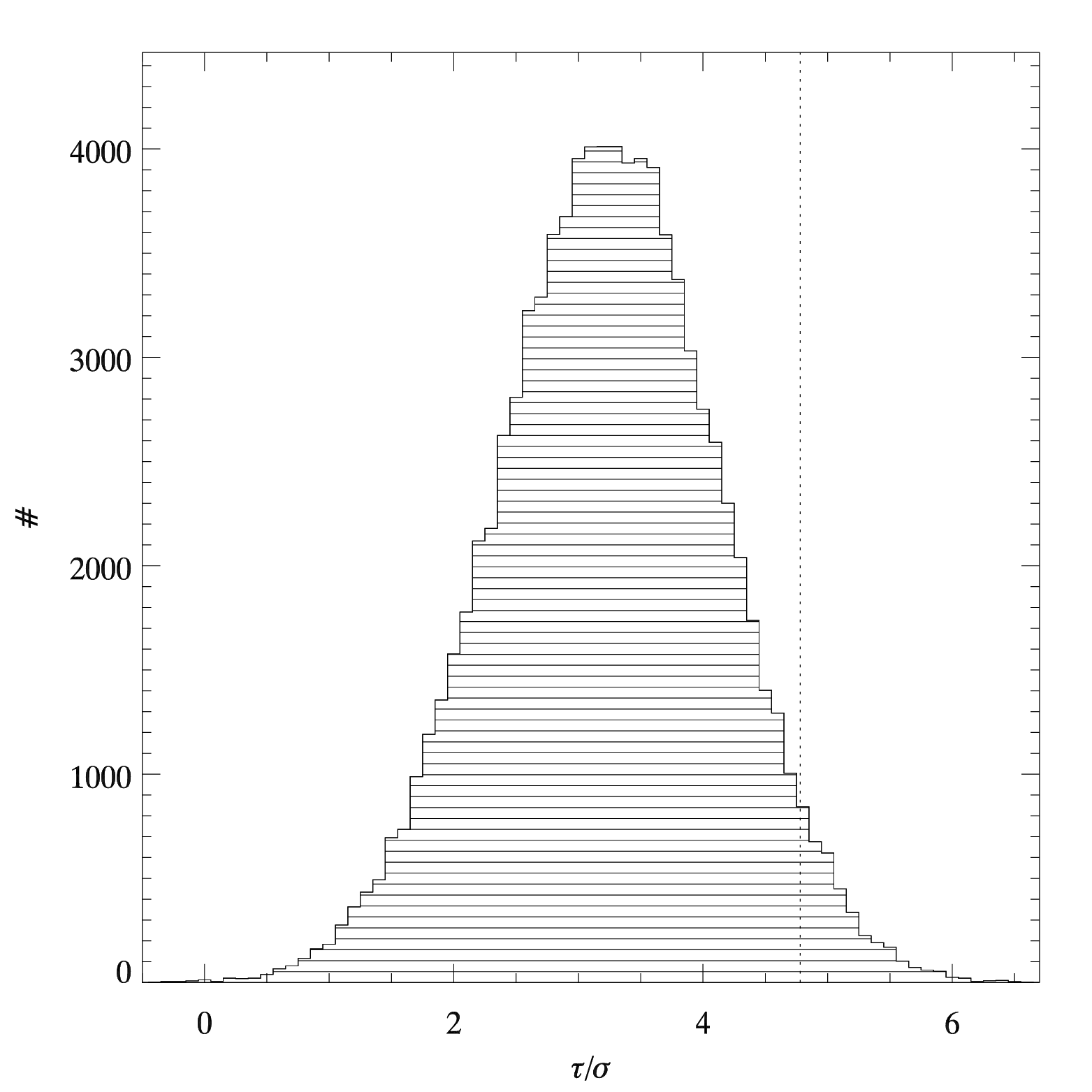, width=9.1cm}
\end{center}
\caption{\label{scrambling}Scrambling test results for the correlations between the hard X-ray luminosity and the radio luminosity at 6 cm (\textit{upper plots}) and 20 cm (\textit{lower plots}). The left panels show the distribution of the Spearman coefficients for the 100\,000 simulated data-sets, in comparison to the one measured for the real data-set (dashed vertical line). Same in the right panels, but for the Kendall partial correlation  coefficient $\tau$ divided by the variance $\sigma$.}
\end{figure*}

We finally performed the ``scrambling test'' \citep[e.g.][]{bregman05,merl06}. The principle at the heart of this test is very simple: we keep together each couple of X-ray luminosity and redshift and ``scramble'' the radio fluxes, assigning each one to a random $L_{2-10}$/z couple. Then, adopting the newly assigned redshift, we calculate the radio luminosity from the random flux and correlate it to the assigned X-ray luminosity. Any physical correlation intrinsic to the original data-set must disappear in the new artificial data-sets, because the radio flux of a given object is now associated to the X-ray luminosity of another one. We produced 100\,000 scrambled data-sets with this procedure for the 6 and the 20 cm radio fluxes and we calculated for each of them the Spearman correlation coefficient, the Kendall partial correlation coefficient and the censored fit. The results are shown in Fig. \ref{scrambling}.

\begin{figure*}
\begin{center}
\epsfig{file=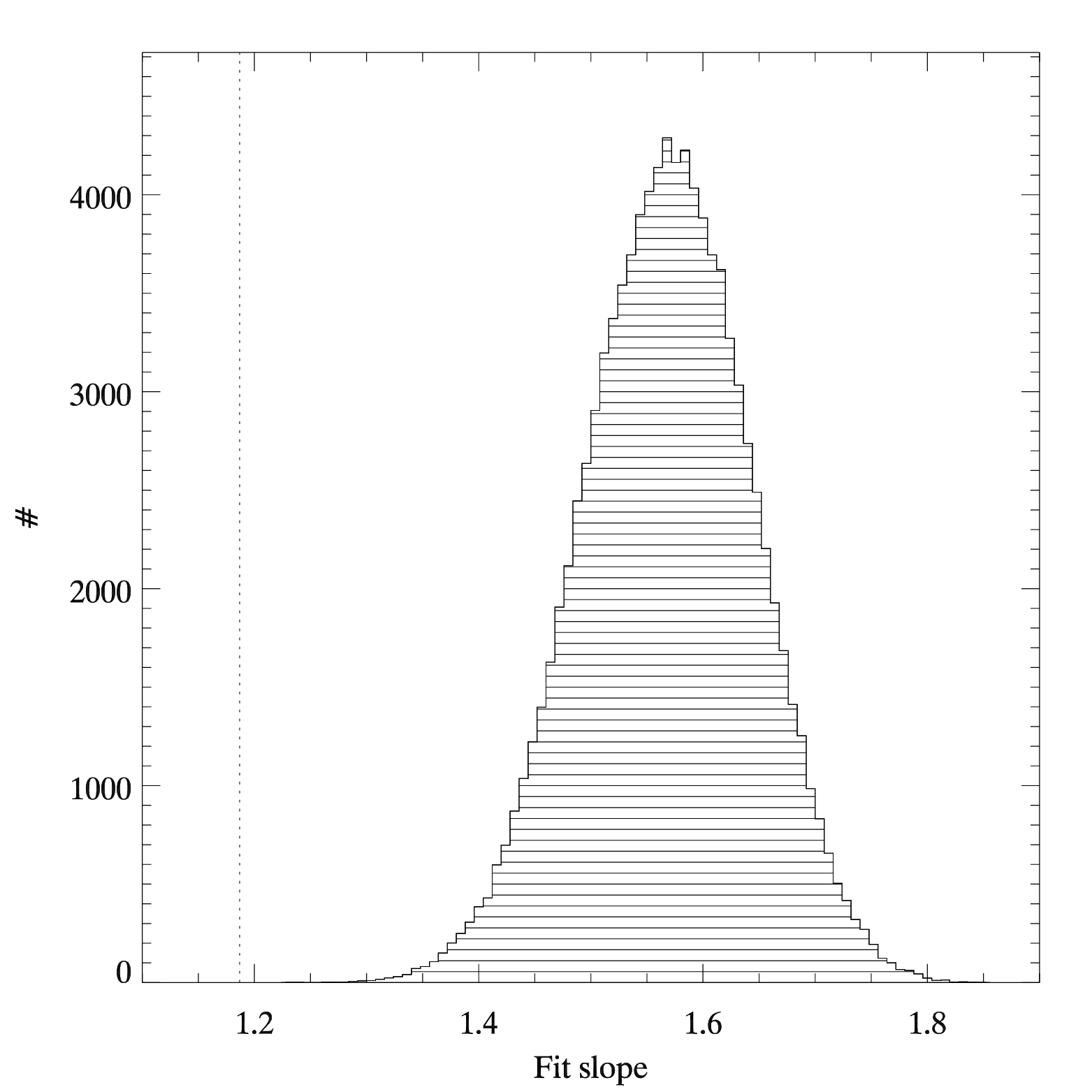, width=9.1cm}
\epsfig{file=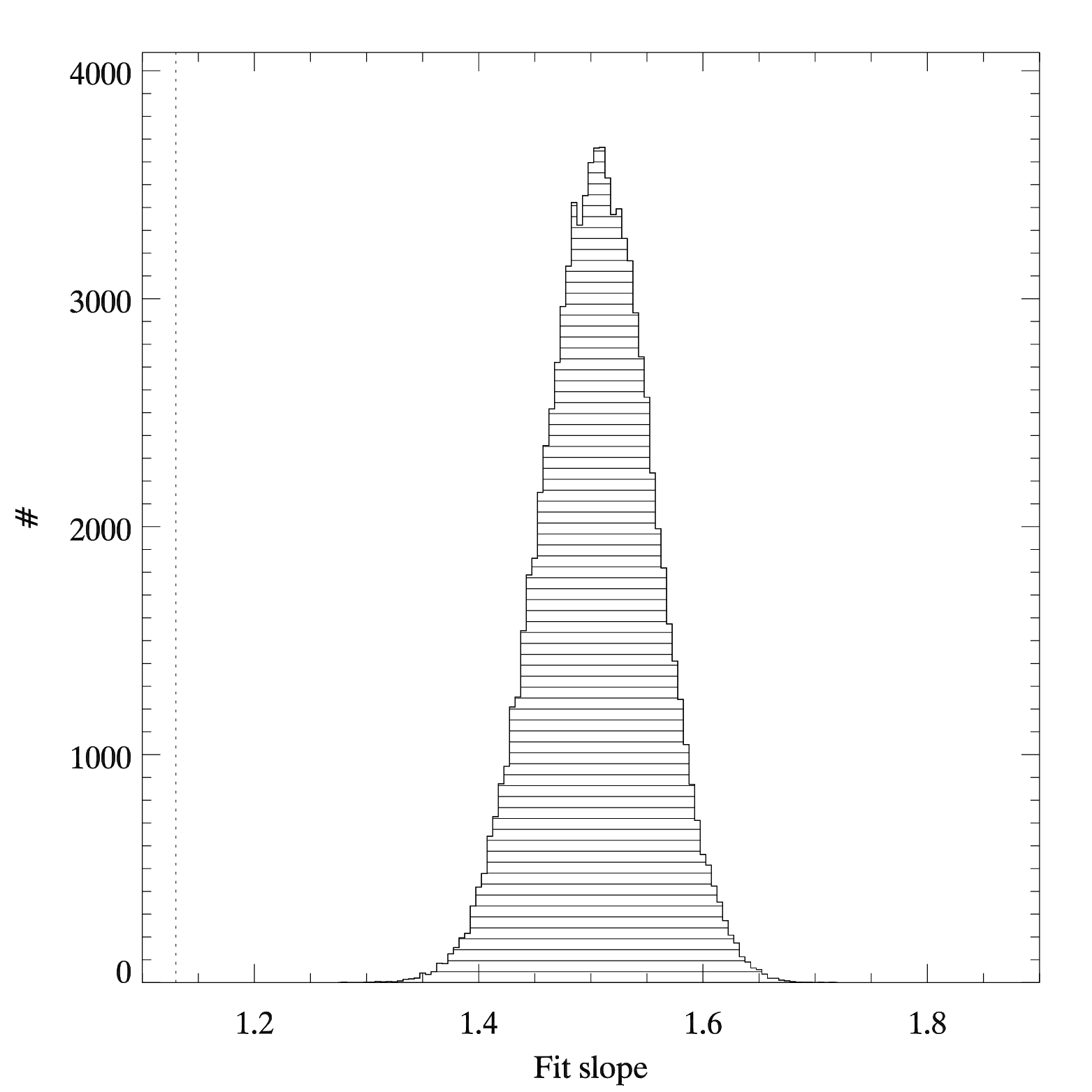, width=9.1cm}
\end{center}
\caption{\label{scrambling_slope}Scrambling test results for the correlations between the hard X-ray luminosity and the radio luminosity at 6 cm (\textit{left}) and 20 cm (\textit{right}). The best fit slope distributions for the 100\,000 simulated data-sets are shown, in comparison to the ones measured for the real data-set (dashed vertical line). See text for details.}
\end{figure*}

It is striking to see how the Spearman correlation coefficients for the real data-sets are well within the distribution of the simulated ones: in the case of the 6 cm data, around 87\,000 values of $\rho$ are actually larger than the `real one'. Therefore, this test suggests that the probability that the original correlation is driven entirely by distance is very high, i.e. 87\%. This probability is much lower for the 20 cm data (which includes upper limits), but still large: 8\%. The simulated distributions of the Kendall partial correlation coefficients are instead shifted towards less significant values with respect to the real ones, but still the probability that distance plays the main role is rather high, being 7 and 4\%, for 6 and 20 cm, respectively. However, in Fig. \ref{scrambling_slope} we also show the distributions of the best fit slopes for the simulated data-sets compared to the ``real'' ones (right plots): the latter are much flatter. This indicates that the observed slopes are not random, and may still reflect a real intrinsic correlation. On the other hand, this effect may also be due to the fact that CAIXA contains few objects with large redshifts and no radio fluxes below a certain limit. These two factors contribute to enhance the population of very luminous radio objects in the simulated data-sets (when an average or large radio flux is associated to a large z), while the region with high redshift and low radio luminosity is still under-populated, leading to steeper slopes, i.e. objects more radio-luminous than the real ones.

\begin{figure}
\begin{center}
\epsfig{file=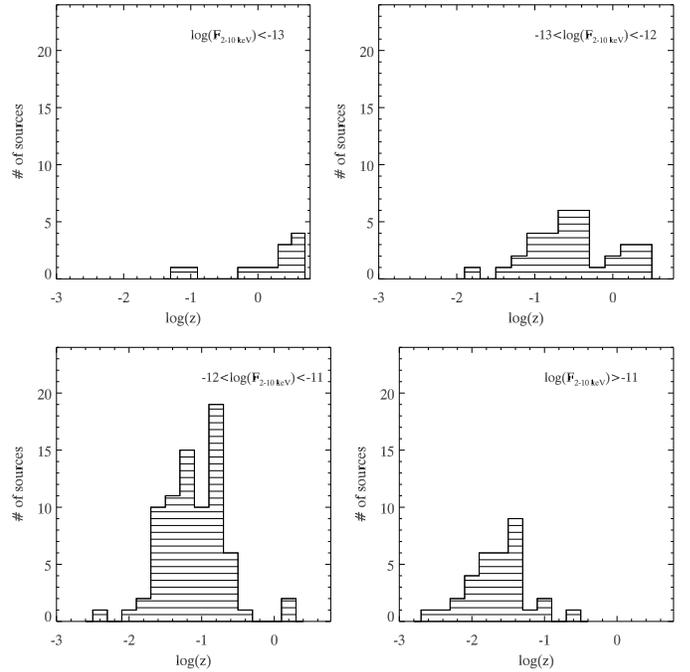, width=9.1cm}
\end{center}
\caption{\label{zdist_fluxbin}Redshift distributions of the sources in CAIXA as a function of hard X-ray flux. The distributions clearly evolve with flux, revealing an intrinsic bias of the catalogue, due to selection effects. See text for details.}
\end{figure}

\begin{figure*}
\begin{center}
\epsfig{file=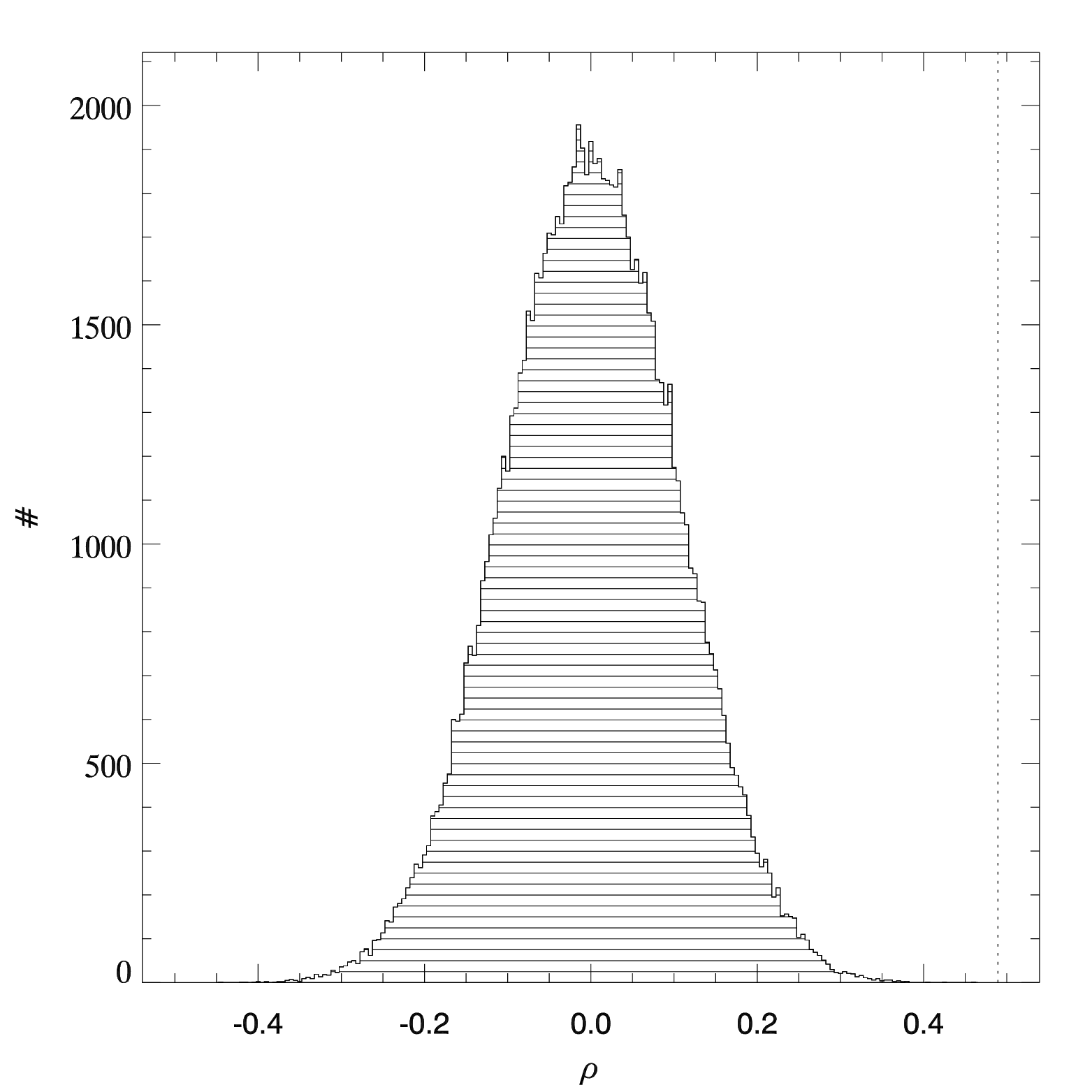, width=9.1cm}
\epsfig{file=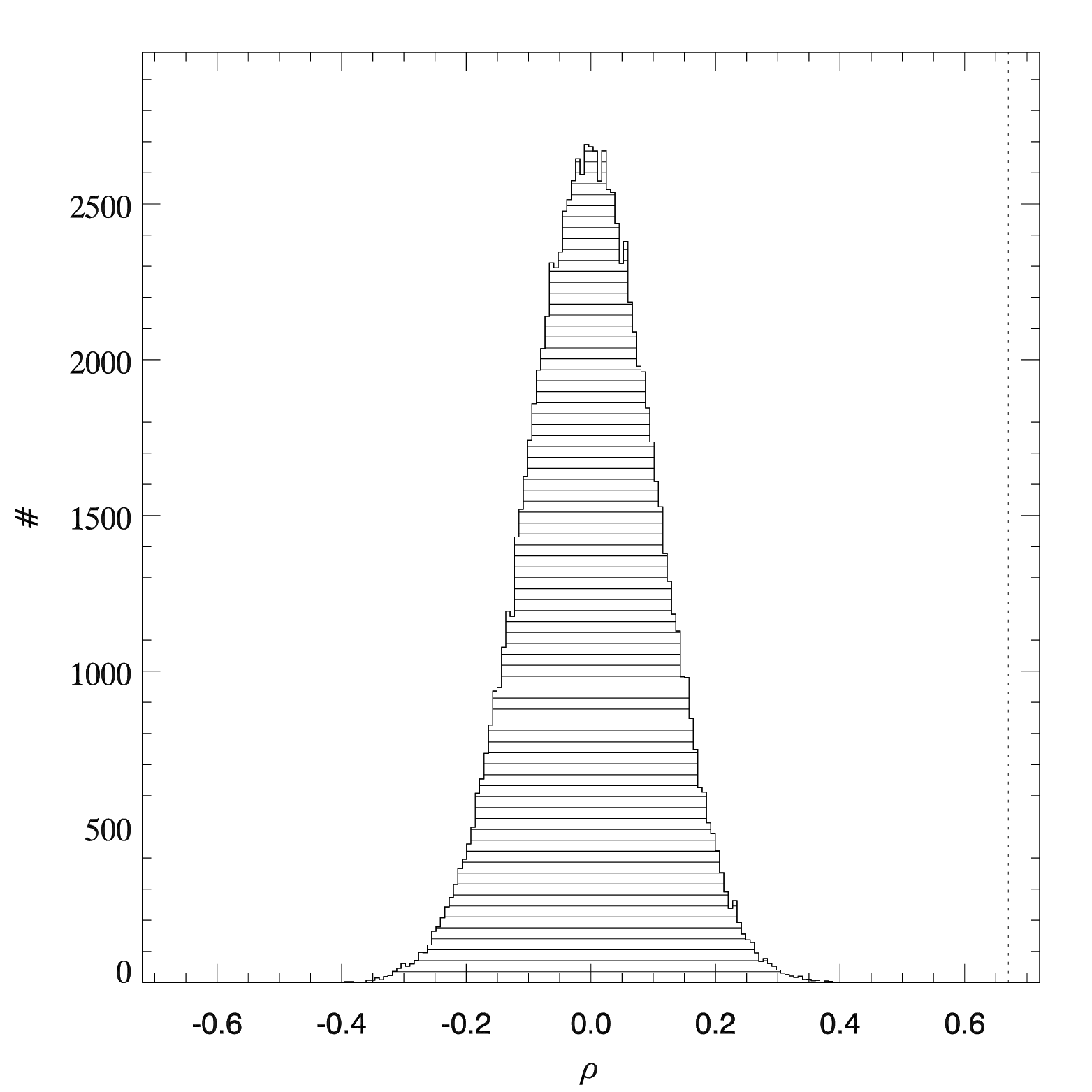, width=9.1cm}
\end{center}
\caption{\label{scrambling_flux}Scrambling test results for the correlations between the hard X-ray flux and the radio flux at 6 cm (\textit{left}) and 20 cm (\textit{right}). The distributions of the Spearman coefficients for the 100\,000 simulated data-sets are shown, in comparison to the ones measured for the real data-set (dashed vertical line).}
\end{figure*}

As a last check, we also performed the ``scrambling test'' for the flux-flux correlations. In this case, the simulated Spearman correlation coefficients are centered (both for 6 and 20 cm) in 0, as expected for random uncorrelated points (see Fig~\ref{scrambling_flux}). Not even a simulated data-set has a correlation coefficient as large as the ones measured in the real data-sets, in agreement with the very low estimated NHP in the latter. However, the strong correlation found between the X-ray and radio fluxes in CAIXA may still be implicitly due by distance effects. Due to selection effects, X-ray fluxes are tightly linked to redshift. This is clearly shown in Fig.~\ref{zdist_fluxbin}, where we plot the redshift distributions of the sources in CAIXA in four hard X-ray flux bins. The evolution of the distributions is evident: faint objects are clustered at high z, while bright sources are peaked at low z. Therefore, although we cannot exclude the presence of a physical correlation between hard X-ray and radio fluxes in CAIXA, we are unable to disentangle it from the selection effects present in the catalogue.

In conclusions, the observed correlation between radio and hard X-ray luminosity appears extremely significant on the basis of standard statistical tests (Spearman correlation coefficients, partial Kendall correlation), but the role of distance effect may be dominant in determining the significance and functional relation. In particular, the scrambling test shows that even the widely used partial Kendall correlation may underestimate the distance effects, thus care must be taken when interpreting the results derived from it. However, random scrambling cannot reproduce the observed slope of the luminosity correlation, and destroys the flux-flux correlation, which is instead observed to be significant. Both results may represent the revealing signs of a real underlying correlation, but they can also be induced by the incompleteness of CAIXA.

Therefore, although a physical correlation may be present in our data, we conservatively cannot draw any conclusion from the X-ray/radio correlations in CAIXA. We would like to stress that this is true for CAIXA, which includes upper limits, but it is by no means a flux-limited or volume-limited sample, while distance effects may of course play less important roles in other catalogues or in complete samples.

\begin{figure*}
\begin{center}
\epsfig{file=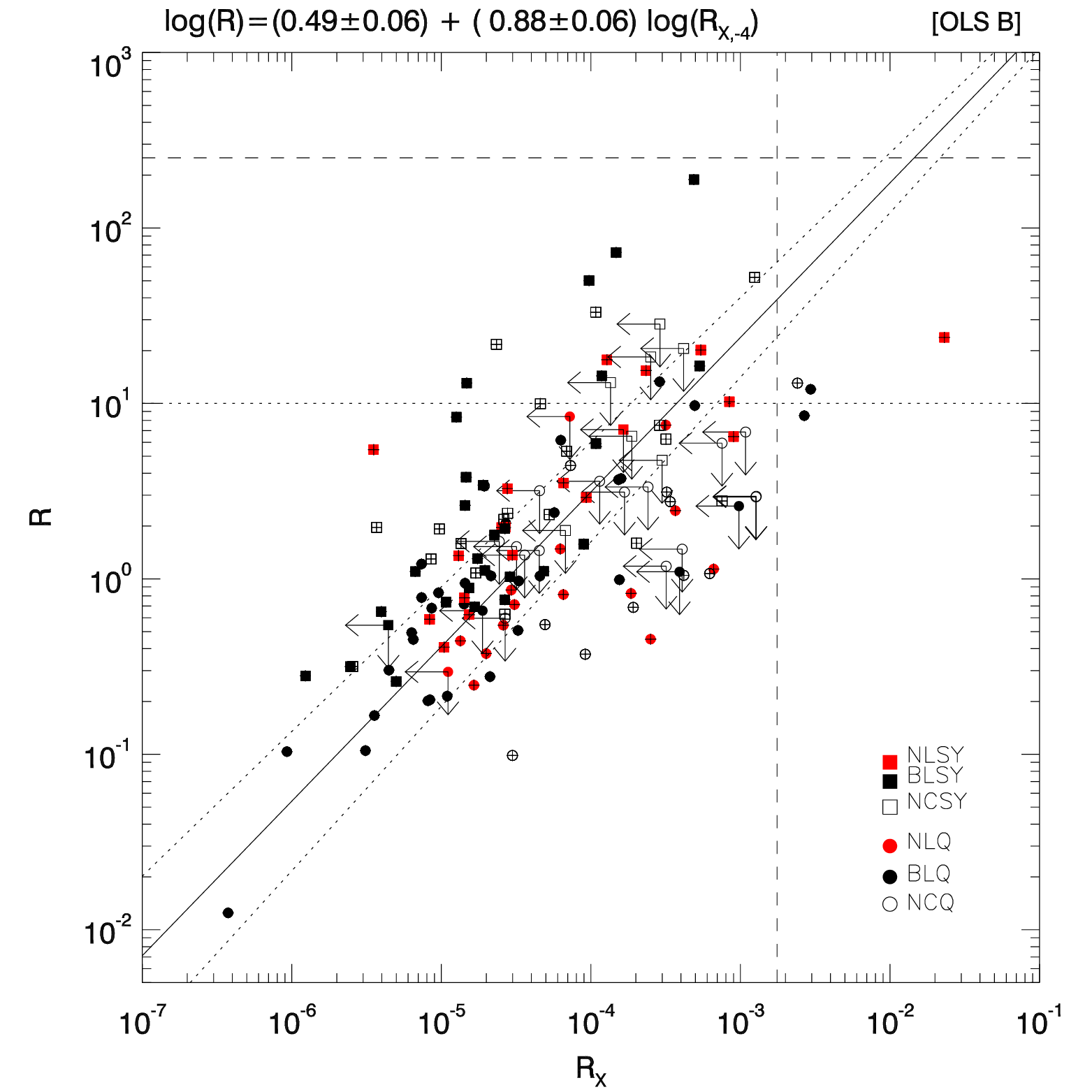, width=9.1cm}
\epsfig{file=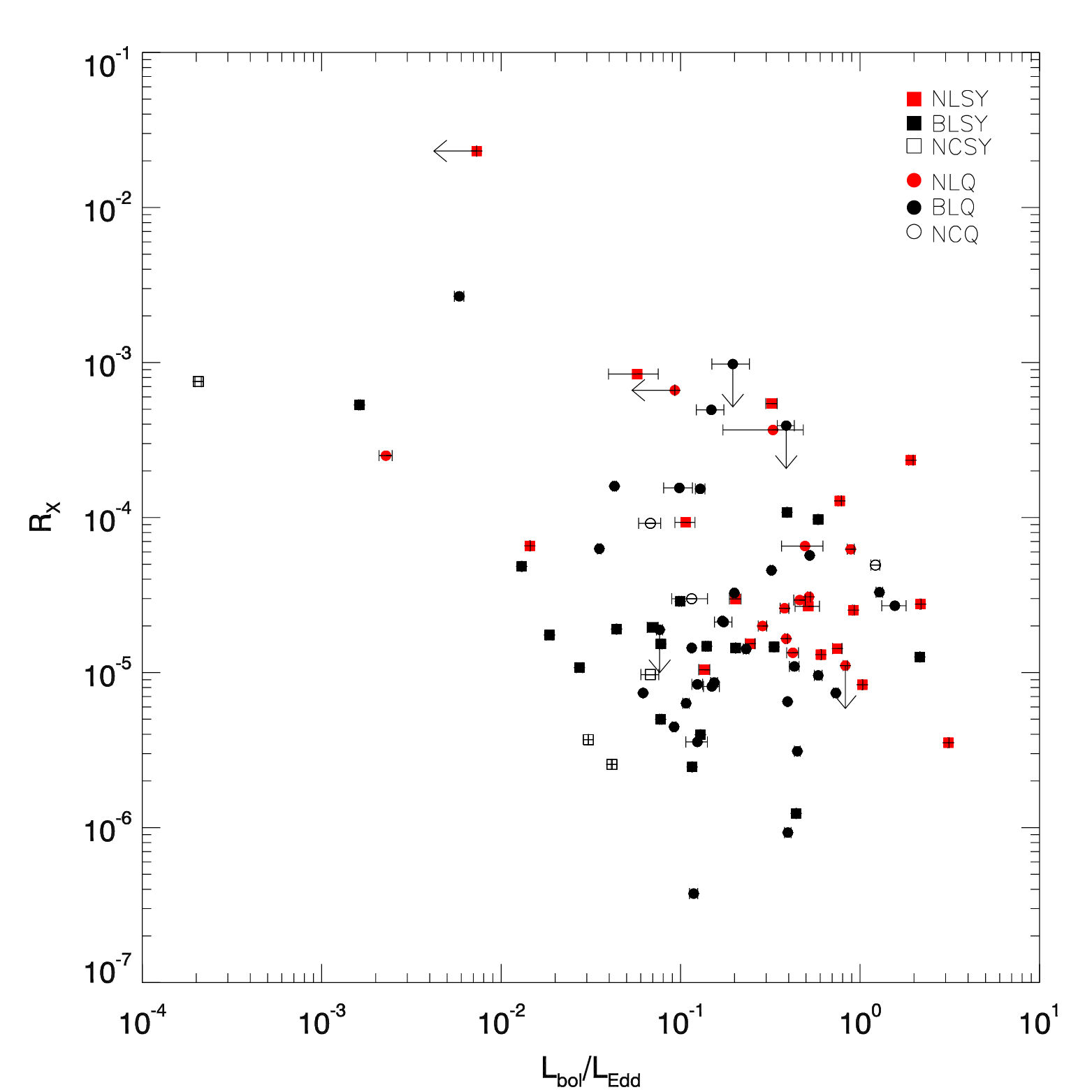, width=9.1cm}
\end{center}
\caption{\label{rx_ro}\textit{Left}: Radio-loudness parameter vs. X-ray radio-loudness parameter (see text for definitions) for all the sources in the catalogue with a radio measure. The dotted horizontal line represents the canonical separation between radio-loud and radio-quiet quasars, while the dashed lines the separation for Seyferts, based both on the optical as well as on the X-ray radio-loudness parameter \citep[after][]{panessa07}. A linear regression fit is superimposed on the plot, along with the best fit parameters. We refer the reader to B09 for the detailed selection criteria adopted in CAIXA with respect to the radio-loudness. \textit{Right}: X-ray radio-loudness parameter against Eddington ratio. Although a trend is observed, the anti-correlation is not statistically significant. See caption of Fig. \ref{lx_lradio} for details on the adopted symbols.}
\end{figure*}

In the left panel of Fig. \ref{rx_ro}, we plot the correlation between the two radio-loudness parameters used to exclude the radio-loud objects in CAIXA, one based on the optical fluxes found in literature, the other on the X-ray fluxes measured in this work (see B09 for details). As expected, a highly significant linear regression between the two parameters is found (see Table \ref{corr}). The slope of the best fit ($0.88\pm0.06$) is perfectly in agreement with the ones found by \citet{tw03} and \citet{panessa07}, even if their samples include also radio-loud objects, thus spanning a larger range both for $\mathrm{R_X}$ and $\mathrm{R_o}$. 

In the right panel of Fig. \ref{rx_ro}, we show the X-ray radio-loudness parameter versus the Eddington ratio. \citet{ho02} found an highly significant anti-correlation between these two parameters, suggesting that weakly active nuclei are powered by advection-dominated accretion flows (ADAFs), which give rise to radio-loud spectra. Although this trend seems to be observed in CAIXA, the anti-correlation is not significant ($\rho=-0.14$ - NHP=0.19). However, this is not unexpected, because our catalogue does not include radio-loud objects and, indeed, the large $\mathrm{R_X}$-low $\mathrm{L_{bol}/L_{Edd}}$ region is not populated in our plot. Finally, we note here that no correlation is apparent between the soft to hard X-ray luminosity ratio and the radio luminosity.

\subsection{\label{gamma}The X-ray spectral index}

Apart from the correlations with H$\beta$ and the X-ray luminosity ratio, already discussed in Sect. \ref{hbeta}, there are no other significant correlations with the hard X-ray spectral index in CAIXA. In particular, in the literature there are claims of a correlation between $\Gamma_h$ and the 2-10 keV luminosity \citep[e.g.][]{dai04,saez08}, but there are also several studies that do not find such a correlation \citep[e.g.][]{rt00,george00,win09}. In CAIXA, this correlation is not significant (96\% confidence level): we note here that this confidence level is not far from the ones quoted by \citet{dai04} and \citet{saez08}, unless the correlation is calculated in particular bins of (relatively high) redshift. In other words, the low significance of the $\Gamma_h$/$L_\mathrm{2-10}$ correlation in CAIXA is in agreement with previous studies, even with those which claim the correlation is real.  Nonetheless, we remind that significant difference between the spectral index of Seyferts and quasars is measured in CAIXA ($\Gamma_{Sy}=1.66\pm0.05$; $\Gamma_{QSO}=1.80\pm0.05$: B09). We showed in B09 that the introduction of a luminosity-dependent Compton reflection component to the baseline model does not remove the difference between these two populations. Further tests should be performed to understand the real driver for this difference. 

Similarly, we do not find a significant correlation between the photon index and the BH mass or the Eddington rate, but the significances found by \citet{pico05} are lower than our threshold, and the results recently found by \citet{kelly08} are strongly dependent on how the BH masses are calculated. On the other hand, the correlation with the Eddington ratio found by \citet{shem08} is more significant than our threshold, but only when high-z and high-luminosity quasars are taken into account.

\subsection{\label{iron}The iron lines}

\citet{bianchi07} already discussed in detail the strong anti-correlations between the Equivalent Width (EW) of the neutral iron narrow emission line and the 2-10 keV luminosity and Eddington rate. Despite the exclusion of a source (PMNJ0623-6436) in CAIXA, the results published in \citet{bianchi07} do not suffer any variation. Recently, \citet{dad08} confirmed the so-called `Iwasawa-Taniguchi' (IT) effect considering the 20-100 keV luminosity, with a slope fully consistent with the one found with CAIXA in the 2-10 keV band. 

It is interesting to note that the likely interpretation of the IT effect as due to the covering factor of the torus decreasing with luminosity would imply a decrease of the fraction of Compton-thick objects with luminosity. This is indeed what found by \citet{dceca08}. Once extracted from their results the fraction of Compton-thick objects with respect to the total AGN population, which can be considered a proxy for the covering factor of the torus, we found that it decreases with luminosity following a slope of $\simeq-0.22$, in very good agreement with the IT effect.

We note here that the IT effect is also in agreement with the recent discovery of a non-linear relation between thermal emission from dust and optical luminosity in AGN, which implies a decrease of the covering factor of dust with luminosity, with a slope of $\simeq-0.18$ \citep{mai07}, i.e. very close to the one which characterises the IT effect. Although the agreement is very good and deserves further investigation, it must be noted that the two methods sample, in principle, different materials. The iron K$\alpha$ line is mainly produced in Compton-thick gas, while the dust can be associated both to Compton-thin and thick gas.

As a final note, we do not find any significant correlation involving the EWs of the ionised iron lines.

\section{Conclusions}

In this paper, we investigated the correlations between the X-ray and the multiwavelength properties of the sources in CAIXA, a Catalogue of AGN in the XMM-\textit{Newton} Archive, presented in detail in a companion paper (B09). The main results can be summarised as follows:

\begin{itemize}

\item \textit{Correlations with H$\beta$ FWHM.} A very strong anti-correlation between the FWHM of H$\beta$ and the ratio between the soft and the hard X-ray luminosity is present in CAIXA. In particular, our catalogue does not contain narrow-line objects with a $L_{0.5-2}/L_{2-10}$ ratio lower than 1. Although this correlation reflects the same physical driver as correlations already published in the literature between the soft X-ray slope and the H$\beta$ FWHM \citep{bbf96,laor97}, we consider that it bears a more fundamental meaning, because it links two more model-independent quantities (it is interesting to notice that no correlation between the soft X-ray slope and the H$\beta$ FWHM is found in CAIXA). The ratio between the optical flux in the V band and the hard X-ray flux is also significantly anti-correlated with the H$\beta$ line width, suggesting that objects with narrower H$\beta$ are X-ray weaker than normal broad-line sources.

\item \textit{Correlations with the BH mass.} The strong correlation between the X-ray luminosity and the BH mass has a slope flatter than 1, suggesting that high-luminosity objects may be X-ray weaker. A luminosity dependent bolometric factor could explain this result, but the one proposed by \citet{mar04} seems too strong at high luminosities. On the other hand, with the bolometric factor proposed by \citet{vf09}, which depends on the Eddington ratio, a relation with a slope consistent with 1 is recovered. In any case, that the intrinsic total power of AGN indeed scales linearly with BH mass is suggested by the linear relation between the radio luminosity and the BH mass. 

\item \textit{Correlations with radio emission.} In CAIXA there is a very strong correlation between the radio (6 or 20 cm) and the X-ray (soft or hard) luminosity. Similar correlations were presented and discussed by several authors in the past \citep[e.g.][]{brink00,panessa07,lww08}. We have critically reviewed the effect that the incompleteness of the sample may have on this correlation. The ``scrambling test" suggests that indeed a distance bias may be dominant, even if the partial Kendall $\tau$ test often used in the literature to account for this bias suggests otherwise. Our conclusion points to the need of using complete unbiased samples, such as that recently discussed by \citet{lb08}, to draw observational constraints on the nature of radio emission in radio-quiet AGN.

\end{itemize}

\acknowledgement
SB and GM acknowledge financial support from ASI (grant I/088/06/0), GP from ANR (ANR-06-JCJC-0047). We acknowledge support
from the Faculty of the European Space Astronomy Centre (ESAC). We thank Andrea Merloni, Eric Feigelson, F. Massaro and G. Miniutti for useful discussions, and Craig Gordon for all his efforts in solving problems related to \textsc{Xspec}. We thank the anonymous referee for valuable suggestions. Based on observations obtained with XMM-Newton, an ESA science mission with instruments and contributions directly funded by ESA Member States and NASA.

\begin{appendix}

\section{\label{appa}Linear fits on censored data}

This appendix is very similar to the one presented in \citet{bianchi07}. We include it here for the reader's convenience and to report the inclusion of the scattering error in the overall uncertainty estimated for the fit parameters.

Since we are dealing with a large number of upper limits, we followed \citet{gua06} to perform a linear fit, taking into account both upper limits and errors on the y parameter. Their method is an extension of the regression method on censored data described by \citet{schm85} and \citet{isobe86} and can be summarized as follows. Several Ordinary Least Squares Y vs. X (OLS Y/X) or Bisector (OLS B) fits (each plot reports on the upper right which method was used) was performed on a set of Monte-Carlo simulated data derived from the experimental points according to the following rules: a) each detection was substituted by a random Gaussian distribution, whose mean is the best-fit measurement and whose standard deviation is its statistical uncertainty; b) each upper limit $U$ was substituted by a random uniform distribution in the interval [0,U]. The mean of the slopes and the intercepts derived from the fits of each data set are our `best fit'.
Note that errors and upper limits on the x parameter are not considered in the procedure, although plotted. This does not affect the results significantly, given the fact that they are typically much smaller than the errors on the y. Moreover, the errors on the best fit parameters only take into account the experimental uncertainties, being the dispersion of these parameters within the Monte Carlo simulated data sets. Any intrinsic, physical scattering of the data points is not included. To test the significance of this anti-correlation, we calculated the Spearman's rank coefficient for each Monte Carlo simulated data set: the mean value is taken as the Spearman's rank coefficient of our `best fit'.

\end{appendix}

\bibliographystyle{aa}
\bibliography{sbs}

\end{document}